\documentclass[12pt]{iopart}

\usepackage{graphicx}
\usepackage{dcolumn}
\usepackage{bm}
\expandafter\let\csname equation*\endcsname\relax

\expandafter\let\csname endequation*\endcsname\relax
\usepackage{amsmath}
\usepackage{amssymb}
\usepackage{amsfonts}
\usepackage{subfigure}
\usepackage{esvect}
\usepackage{hyperref}
\usepackage[dvipsnames]{xcolor}
\usepackage{fix-cm} 
\usepackage{cite}
\usepackage{soul}
\usepackage{mdframed}
\usepackage{lipsum}
\usepackage{multicol}
\usepackage{float}

\newcommand{\CL}{{\tt ${\mathcal C}$osmo${\mathcal L}$attice}~}
\newcommand{\CLns}{{\tt ${\mathcal C}$osmo${\mathcal L}$attice}}
\newcommand{\dx}{\ensuremath{\delta x}}
\newcommand{\dt}{\delta t}

\newcommand{\dd}{\text{d}}

\newcommand{\be}{\begin{equation}}
    \newcommand{\ee}{\end{equation}}
\newcommand{\bea}{\begin{eqnarray}}
    \newcommand{\eea}{\end{eqnarray}}

\begin{document}

\title{Present and future of \CL}

\author{Daniel G. Figueroa\,$^1$, Adrien Florio\,$^2$, and Francisco Torrenti\,$^1$\vspace*{0.35cm}\\
    $^1${\it
        Instituto de F\'isica Corpuscular (IFIC), Consejo Superior de Investigaciones Cient\'ificas (CSIC) and Universitat de Val\`encia, 46980, Valencia, Spain.}\\
    $^2${\it Department of Physics and Astronomy, Stony Brook University, New York 11794, USA}\\
}

\ead{daniel.figueroa@ific.uv.es, adrien.florio@stonybrook.edu, f.torrenti@uv.es}
\vspace{10pt}
\begin{indented}
\item[]December 2023
\end{indented}

\begin{abstract}
We discuss the present state and planned updates of \CLns, a cutting-edge code for lattice simulations of non-linear dynamics of scalar-gauge field theories in an expanding background. We first review current capabilities of the code, including the simulation of interacting singlet scalars and of Abelian and non-Abelian scalar-gauge theories. We also comment on new features recently implemented, such as the simulation of gravitational waves from scalar and gauge fields. Secondly, we discuss new extensions of \CL that we plan to release publicly. On the one hand, we comment on new physics modules, which include axion-gauge interactions $\phi F\tilde F$, non-minimal gravitational couplings $\phi^2 R$, creation and evolution of cosmic defect networks, and magneto-hydro-dynamics (MHD). On the other hand, we discuss new technical features, including evolvers for non-canonical interactions, arbitrary initial conditions, simulations in 2+1 dimensions, and higher accuracy  spatial derivatives.
\end{abstract}

%
%
%
%
%

\section{The {\it Numerical} Early Universe} 

One of the cornerstones of modern cosmology is {\it inflation}, basically defined as a period of accelerated expansion in the early universe~\cite{Guth:1980zm, Linde:1981mu, Albrecht:1982wi,Brout:1977ix,Starobinsky:1980te,Kazanas:1980tx,Sato:1980yn}. Introduced to overcome the limitations of the hot Big Bang framework~\cite{Guth:1980zm, Linde:1981mu}, inflation also provides a mechanism to generate the primordial density perturbations~\cite{Mukhanov:1981xt, Guth:1982ec,Starobinsky:1982ee, Hawking:1982cz,Bardeen:1983qw}. The inflationary period is typically thought to be driven by a scalar field, the \textit{inflaton}, with potential chosen to sustain an accelerated expansion for at least $\sim$ 50-60 e-folds. After inflation a period of \textit{reheating} must follow, transferring the energy available into other fields, which eventually dominate the energy budget and thermalize, giving pass to the standard hot Big Bang expansion. Reviews on inflation and reheating are available in \cite{Lyth:1998xn,Riotto:2002yw,Bassett:2005xm,Linde:2007fr,Baumann:2009ds} and~\cite{Allahverdi:2010xz,Amin:2014eta,Lozanov:2019jxc,Allahverdi:2020bys}, respectively.

The phenomenology during the early universe, both during and after inflation, can be very rich and often involves non-linear physics. Non-linear phenomena in the early universe include preheating and other particle production mechanisms~\cite{Traschen:1990sw,Kofman:1994rk,Shtanov:1994ce,Kaiser:1995fb,Kofman:1997yn,Greene:1997fu,Kaiser:1997mp,Kaiser:1997hg,Greene:1998nh,Greene:2000ew,Peloso:2000hy,Berges:2010zv,Enqvist:2012tc}, the generation of scalar metric perturbations~\cite{Bassett:1998wg,Bassett:1999mt,Bassett:1999ta,Finelli:2000ya,Chambers:2007se,Bond:2009xx,Imrith:2019njf,Musoke:2019ima,Giblin:2019nuv,Martin:2020fgl,Adshead:2023mvt} possibly leading to the formation of primordial black holes~\cite{Cotner:2019ykd,Martin:2019nuw,GarciaBellido:1996qt,Green:2000he,Hidalgo:2011fj,Torres-Lomas:2014bua,Suyama:2004mz,Suyama:2006sr,Cotner:2018vug}, phase transitions \cite{Rajantie:2000fd,Hindmarsh:2001vp,Copeland:2002ku,GarciaBellido:2002aj,Niemi:2018asa,Mazumdar:2018dfl,Hindmarsh:2020hop,Brandenburg:2017neh,Brandenburg:2017rnt}, the creation of topological defects \cite{Hindmarsh:1994re,Felder:2000hj,Hindmarsh:2000kd,Rajantie:2001ps,Rajantie:2002dw,Donaire:2004gp,Copeland:2009ga,Dufaux:2010cf,Hiramatsu:2012sc,Kawasaki:2014sqa,Fleury:2016xrz,Moore:2017ond} and their subsequent evolution \cite{Vincent:1997cx,Bevis:2006mj,Hindmarsh:2014rka,Daverio:2015nva,Lizarraga:2016onn,Hindmarsh:2018wkp,Gorghetto:2018myk,Buschmann:2019icd,Hindmarsh:2019csc,Eggemeier:2019khm,Hindmarsh:2021vih,Blanco-Pillado:2023sap}, the creation of soliton-like structures like oscillons and others~\cite{Amin:2011hj,Zhou:2013tsa,Antusch:2016con,Antusch:2017flz,Lozanov:2017hjm,Amin:2018xfe,Liu:2018rrt,Kitajima:2018zco,Lozanov:2019ylm,Antusch:2019qrr,Nazari:2020fmk,Aurrekoetxea:2023jwd}, etc. These non-linear phenomena may have important observable implications such as e.g.~the creation of gravitational waves~\cite{Khlebnikov:1997di,Easther:2006gt,Easther:2006vd,GarciaBellido:2007af,Dufaux:2007pt,Dufaux:2008dn,Dufaux:2010cf,Figueroa:2012kw,Hiramatsu:2013qaa,Hindmarsh:2013xza,Zhou:2013tsa,Bethke:2013aba,Bethke:2013vca,Hindmarsh:2015qta,Antusch:2016con,Hindmarsh:2017gnf,Antusch:2017flz,Antusch:2017vga,Figueroa:2017vfa,Cutting:2018tjt,Liu:2018rrt,Lozanov:2019ylm,Adshead:2019lbr,Adshead:2019igv,Cutting:2019zws,Pol:2019yex,Figueroa:2020lvo,Cutting:2020nla,Figueroa:2022iho,Cosme:2022htl,Klose:2022knn,Cui:2023fbg,Baeza-Ballesteros:2023say,Servant:2023tua} (see \cite{Caprini:2018mtu} for a review), the generation of the dark matter relic abundance \cite{Garcia:2018wtq,Garcia:2021iag,Garcia:2022vwm,Lebedev:2022vwf,Zhang:2023xcd}, the realization of magnetogenesis~\cite{DiazGil:2005qp,DiazGil:2007qx,DiazGil:2007dy,DiazGil:2008tf,Fujita:2016qab,Adshead:2016iae,Vilchinskii:2017qul} and baryogenesis mechanisms~\cite{Kolb:1996jt,Kolb:1998he,GarciaBellido:1999sv,Allahverdi:2000zd,Rajantie:2000nj,Cornwall:2001hq,Copeland:2001qw,Smit:2002yg,GarciaBellido:2003wd,Tranberg:2003gi,Tranberg:2009de,Kamada:2010yz,Lozanov:2014zfa,Adshead:2016iae}, or the determination of the equation of state after inflation and its implications for CMB observables~\cite{Podolsky:2005bw,Lozanov:2016hid,Figueroa:2016wxr,Lozanov:2017hjm,Krajewski:2018moi,Antusch:2020iyq,Saha:2020bis,Antusch:2021aiw,Mansfield:2023sqp}.

In order to make reliable predictions of the above phenomena, we need to capture their non-linear dynamics with appropriate numerical tools. The gradual development of such tools is giving rise to an emergent field, dedicated to address the non-linear dynamics of early universe phenomena. This field, which we like to refer to as $\mathcal{L}${\it attice} $\mathcal{C}${\it osmology}, is gaining more and more attention in the recent years, as reflected by the number of dedicated numerical packages developed during the last two decades, see~\cite{Felder:2000hq,Felder:2007nz,Frolov:2008hy,Sainio:2009hm,Easther:2010qz,Huang:2011gf,Sainio:2012mw,Child:2013ria,Daverio:2015ryl,Amin:2018xfe,Lozanov:2019jff,Giblin:2019nuv,Andrade:2021rbd}. It is in this context that our package \CL was developed.

\CL is a code for real-time simulations in a lattice of scalar-gauge field theory dynamics in an expanding universe. It was created purposely to explore the phenomenology and observational implications of {\it non-linear} dynamics in the early Universe, see \href{http://www.cosmolattice.net}{\color{blue} http://www.cosmolattice.net}. Version 1.0 of the code was publicly released in February 2021, with the capability of simulating the dynamics of interacting canonical scalar theories and $SU(2)$$\times$$U(1)$ gauge theories evolving in a spatially-flat expanding background. The equations are solved by different numerical algorithms with accuracies that range from $\mathcal{O}(\delta t^2)$ to $\mathcal{O}(\delta t^{10})$, and in the case of gauge theories they preserve the Gauss constraint up to machine precision. \CL is written in C++, and uses a modular structure such that the technical details of the code are separated from the physics implementation. These are dealt by the {\tt TempLat} and {\tt CosmoInterface} libraries of the code, respectively. Regarding the {\tt CosmoInterface}, it establishes a unique symbolic language, wherein field variables and their associated operations are defined, enabling the introduction of differential equations and operators in a manner closely resembling the continuum. The code is parallelized with Message Passing Interface (MPI), and also uses a discrete Fourier transform parallelized in multiple spatial dimensions implemented in the PFFT library, which allows to run the code in clusters of thousands of cores with almost perfect scalability. The release of the code was accompanied by an extensive monographic review on lattice techniques~\cite{Figueroa:2020rrl}, which constitutes the theoretical basis of the code. An extensive user manual was also released~\cite{Figueroa:2021yhd}. 

Since the initial release of \CLns, we have updated the code with new features, which we have documented in {\it Technical Notes} made available in our website, see 
\href{http://cosmolattice.net/technicalnotes}{\color{blue}http://cosmolattice.net/technicalnotes}. For example, since version 1.1 (released in May 2022), it is possible to simulate the gravitational waves sourced by scalar singlets. Later on, in version 1.2 (released in June 2023), we incorporated the possibility of simulating gravitational waves sourced by a $U(1)$ gauge sector. We are currently implementing new physics and technical features in the code, which we plan to release in following updates of the code. The aim of this manuscript is precisely to review the current capabilities of \CLns, while providing an outlook of future upgrades on physics interactions and technical capabilities that we plan to add.

The structure of this document is as follows. In Sect.~\ref{sec:present} we review the physics that \CL can currently simulate. In Sect.~\ref{sec:futurep1} we present new physics that will be incorporated in the foreseeable future, whereas in Sect.~\ref{sec:futurep2} we review upcoming technical features. In Sect.~\ref{sec:outlook} we provide a final outlook. We use natural units $c=\hbar=1$ and choose the metric signature $(-1,+1,+1,+1)$. The reduced Planck mass is $m_p \simeq 2.44\cdot 10^{18}$ GeV.

\section{Present capabilities of \CL}  \label{sec:present}

\CL evolves the equations of motion (EOM) of the interacting fields in a regular cubic lattice of comoving side length $L$ with $N$ sites per dimension. The smallest distance we can probe in a lattice is the {\it lattice spacing} $\delta x \equiv L/ N$. A finite range of discrete momenta is captured from a minimum infrared scale $k_{\rm IR} = 2\pi / L$ to a maximum ultraviolet cut-off $k_{\rm UV} = \sqrt{3}Nk_{\rm IR}/2 = \sqrt{3}\pi/\delta x$. 

More specifically, \CL solves a discretized version of the fields' EOM in which continuous derivatives are replaced by discretized versions that approximate them up to a certain accuracy order in the lattice spacing. These equations are solved through evolution algorithms with appropriately chosen time step $\delta t$. Examples of algorithms include \textit{staggered leapfrog}, \textit{velocity/position verlet}, \textit{Runge-Kutta}, \textit{Yoshida}, and \textit{Gauss-Legendre}, see Ref.~\cite{Figueroa:2020rrl} for an extensive discussion. The current version of \CL implements two families of evolution algorithms: staggered leapfrog of accuracy $\mathcal{O}(\dt^2)$, and velocity verlet with degrees of accuracy ranging from $\mathcal{O}(\delta t^2)$ to $\mathcal{O}(\delta t^{10})$. We plan to include other evolution algorithms such as \textit{Runge-Kutta} in the near future, which are necessary to solve the dynamics of e.g.~non-canonical interactions, see Sect.~\ref{sec:NonCanonicalDynamics}. We also highlight that  \CL can produce  three different kinds of output at any time during the evolution: \textit{volume averages} (of e.g.~field amplitudes or energy components), \textit{field spectra}, and \textit{snapshots} (i.e.~3-dimensional distributions).

In the following we describe the dynamics that the current version of \CL can solve (Sect.~\ref{subsec:expansion} to \ref{subsec:GWs}), how the initial conditions of the different fields are set (Sect.~\ref{subsec:initcond}), and the kinds of outputs that \CL can generate (Sect.~\ref{sec:output}). We note that the different matter sectors described in Sect.~\ref{sec:scalar}, \ref{subsec:Abelian} and \ref{subsec:NonAbelian} (scalar, Abelian gauge and non-Abelian gauge, respectively) are discussed separately for pedagogical purposes, but they can be activated simultaneously. 

\subsection{Expanding background} \label{subsec:expansion}

\CL solves the field dynamics on a flat {\it Friedmann-Lema\^itre-Robertson-Walker} (FLRW) metric, described by the line element,
\begin{eqnarray}
\dd s^2 = g_{\mu\nu}\dd x^\mu\dd x^\nu = - a^{2 \alpha}(\eta) \dd \eta^2 + a^2 (\eta) \delta_{ij} \dd x^i \dd x^j \ , \label{eq:FLRWmetric}
\end{eqnarray}
where $a(\eta)$ is the scale factor and $\eta$ is the so-called {\it $\alpha$-time variable}, characterized by the choice of a constant parameter $\alpha$. For example, $\eta$ is cosmic time for $\alpha = 0$ and conformal time for $\alpha = 1$. \CL can solve the field EOM for any reasonable choice of $\alpha$. Time-derivatives with respect to $\alpha$-time will be denoted as $' \equiv d /d \eta$.

By particularizing the Einstein's field equations to the metric \eqref{eq:FLRWmetric}, we obtain the 1st and 2nd Friedmann equations,
\begin{eqnarray}
\mathcal{H}^2 \equiv  \left({a'\over a}\right)^2 =  \frac{a^{2 \alpha}}{3 m_p^2}\left\langle \rho \right\rangle \,, \hspace{0.7cm}
{a''\over a} = \frac{a^{2 \alpha}}{6 m_p^2}\left\langle (2 \alpha-1)  {\rho} - 3  p \right\rangle \,,  \label{eq:sfEOM}
\end{eqnarray}
where $\rho$ and $p$ are the energy and pressure densities of the fields sourcing the expansion, and $\langle \dots \rangle$ is a volume average over the entire lattice. \CL solves the 2nd Friedmann equation together with the fields' EOM (see below) in a \textit{self-consistent} manner, so that the evolving fields act as a source in~\eqref{eq:sfEOM}. To check for the accuracy of the evolution, the code uses the 1st Friedmann equation, which is only obeyed to a certain accuracy in the lattice, typically to better than $\mathcal{O}(0.1)\%$, depending on the time evolver. Alternatively, \CL can assume a \textit{fixed background} scenario, in which the expansion is sourced by an unspecified energetically-dominant fluid with a given equation of state $w \equiv \left\langle p \right\rangle / \left\langle \rho \right\rangle = {\rm const}$ [e.g.~$w=1/3$ for radiation domination ({\rm RD}) or  $w=0$ for matter domination ({\rm MD})].

\subsection{Canonical scalar theories} \label{sec:scalar}

\CL can simulate canonical scalar theories based on $N_s$ interacting scalar fields $\lbrace \phi_{a} \rbrace$, $a=1,\dots , N_s$, described by the action
\begin{eqnarray}
S = - \int d^4x\, \sqrt{-g}\left(\frac{1}{2} \sum_{b=1}^{N_s} \partial_{\mu} \phi_b \partial^{\mu} \phi_b + V(\lbrace \phi_a \rbrace) \right) \ . 
\end{eqnarray}
Here, $V (\{\phi_a\})$ is the potential describing the interactions between the fields. By minimizing the action we obtain the scalar field EOM,
\begin{eqnarray}
\phi_a'' - a^{-2 (1  - \alpha )} \nabla^2 \phi_a + (3 - \alpha)\mathcal{H} \phi'_a +  a^{2 \alpha} V_{,\phi_a} = 0\,. \label{eq:EOMscalarContinuumNat}
\end{eqnarray}
The Friedmann's equations \eqref{eq:sfEOM} specialize in this case to 
\begin{eqnarray}
\hspace*{0.75cm} {a''\over a} = \frac{a^{2 \alpha}}{3 m_p^2}\big\langle (\alpha-2){K}_{\phi} + \alpha {G}_{\phi}  + (\alpha + 1)V \big\rangle \,,
\label{eq:friedmanS2}\\
\left({a'\over a}\right)^2 =  \frac{a^{2 \alpha}}{3 m_p^2}\big\langle  {K}_{\phi} + {G}_{\phi} + {V}\big\rangle \,,
\label{eq:friedmanS1}
\end{eqnarray}
where we have used that the energy and pressure densities of the scalar fields are
\begin{eqnarray}
    \rho = {K}_{\phi} + {G}_{\phi}  + {V} \ , \hspace{0.5cm} p = {K}_{\phi} - {1\over3}{G}_{\phi} - {V} \ ,  \label{eq:pLocal}
\end{eqnarray}
with ${K}_{\phi} \equiv  \sum_b {\phi'_b}^2 /(2a^{2\alpha})$ and ${G}_{\phi} \equiv  \sum_{i,b} (\partial_i \phi_b)^2 / (2a^{2})$ the kinetic and gradient contributions. \CL can solve the scalar EOM~(\ref{eq:EOMscalarContinuumNat}), together with Eq.~(\ref{eq:friedmanS2}) for the expansion of the Universe. We use the Friedmann equation (\ref{eq:friedmanS1}) as a consistency check of ``energy conservation". The \CL module for canonical scalar field theories has been extensively applied to the study of different models of (p)reheating after inflation, see e.g.~\cite{Antusch:2020iyq,Antusch:2021aiw,Antusch:2022mqv,Dux:2022kuk,Garcia:2023eol,Garcia:2023dyf,Matsui:2023wxm,Matsui:2023ezh,Mansfield:2023sqp}.

\subsection{Abelian gauge theories}
\label{subsec:Abelian}

\CL can also simulate Abelian gauge fields coupled to charged scalar fields, described by the action
\begin{eqnarray}
S = - \int d^4 x\sqrt{-g} \left((D_{\mu}^A \varphi)^{*}(D_A^{\mu} \varphi)  + \frac{1}{4} F_{\mu \nu} F^{\mu \nu} +  V(\varphi)\right)\,,
\label{eq:Lagrangian}
\end{eqnarray}
with $V(\varphi) \equiv V(|\varphi|)$ a potential term,
$\varphi \equiv {1\over\sqrt{2}}(\varphi_0 +i\varphi_1)  
$ a charged scalar field, and where we have introduced the standard {\it covariant derivative} (denoting $Q_{A}$ the Abelian charge of the scalar field) and {\it field strength} tensor as
\bea
D_{\mu}^{\rm A} & \equiv &  \partial _{\mu} - i  g_AQ_AA_\mu \ ,
\label{eq:AbCovDerivCont}\\
F_{\mu \nu} &\equiv & \partial_{\mu}  A_{\nu} - \partial_{\nu} A_{\mu} \ . \label{eq:FmnAbelian}
\eea
Varying action~(\ref{eq:Lagrangian}) leads to the following EOM,
\begin{eqnarray}
\label{eq:higgsU1-eom}
\varphi'' - a^{-2(1 - \alpha)} {\vv D}_{\hspace{-0.5mm}A}^{\,2}\varphi + (3 - \alpha)\frac{{a'}}{a}  {\varphi'} = - \frac{a^{2 \alpha}V_{,|\varphi|} }{2} \frac{\varphi}{|\varphi |} \ ,\\
\label{eq:U1eom}
\partial_0 F_{0i} - a^{-2(1 - \alpha )}\partial_j F_{ji} + (1 - \alpha) \frac{{a'}}{a} F_{0i} =
a^{2 \alpha}J^A_i \ , 
\vspace*{1cm}\\
\label{eq:GaussU1-eom}
\partial_i F_{0i} = a^2J^A_0 \ , 
\end{eqnarray}
where we have introduced the Abelian charge current
\begin{eqnarray}
\label{eq:AbelianCurrent}
J_A^\mu \equiv 2g_AQ_A^{(\varphi)} \mathcal{I}m [ \varphi^{*} ( D_A^{\mu} \varphi )] \,.\end{eqnarray}
The Friedmann's equations \eqref{eq:sfEOM} specialize in this case to 
\begin{eqnarray}
{a''\over a} = \frac{a^{2 \alpha}}{3 m_p^2}\big\langle (\alpha-2){K}_{\varphi} + \alpha {G}_{\varphi}  + (\alpha + 1)V + (\alpha-1)({K}_{U(1)} + {G}_{U(1)}) \big\rangle \,,
\label{eq:friedmanU2}\\
\hspace*{-0.75cm} \left({a'\over a}\right)^2 =  \frac{a^{2 \alpha}}{3 m_p^2}\big\langle  {K}_{\varphi} + {G}_{\varphi}+ {K}_{U(1)} + {G}_{U(1)} + {V}\big\rangle \,,
\label{eq:friedmanU1}
\end{eqnarray}
with
\begin{eqnarray}
\label{eq:phiE}
{K}_{\varphi} = \frac{1}{a^{2\alpha} } (D_0^A \varphi)^*(D_0^A \varphi)\,, \hspace*{1cm} {G}_{\varphi} = \frac{1}{a^2} \sum_i (D_i^A \varphi)^*(D_i^A \varphi)\,,
\\
\label{eq:gaugeE}
{K}_{U(1)} = \frac{1}{2 a^{2 + 2 \alpha}}  \sum_{i} F_{0i}^2 \ , \hspace*{1cm} {G}_{U(1)} = \frac{1}{2 a^4}  \sum_{i,j<i} F_{ij}^2\,,
\end{eqnarray}
the kinetic and gradient energies of the charged scalar [Eq.~(\ref{eq:phiE})] and gauge [Eq.~(\ref{eq:gaugeE})] sectors.

\CL can solve the EOM~(\ref{eq:higgsU1-eom})-(\ref{eq:U1eom}) together with Eq.~(\ref{eq:friedmanU2}) for the expansion of the Universe. Note that Eq.~(\ref{eq:GaussU1-eom}) is the Gauss constraint, which is a consequence of gauge invariance. As such, it needs to be preserved up to machine precision during time evolution, something taken care of by the algorithms implemented in \CLns. We use the Friedmann equation (\ref{eq:friedmanU1}) as a consistency check of ``energy conservation". 

\subsection{Non-Abelian gauge theories}
\label{subsec:NonAbelian}

\begin{figure}
    \centering
    \includegraphics[width=.63\textwidth]{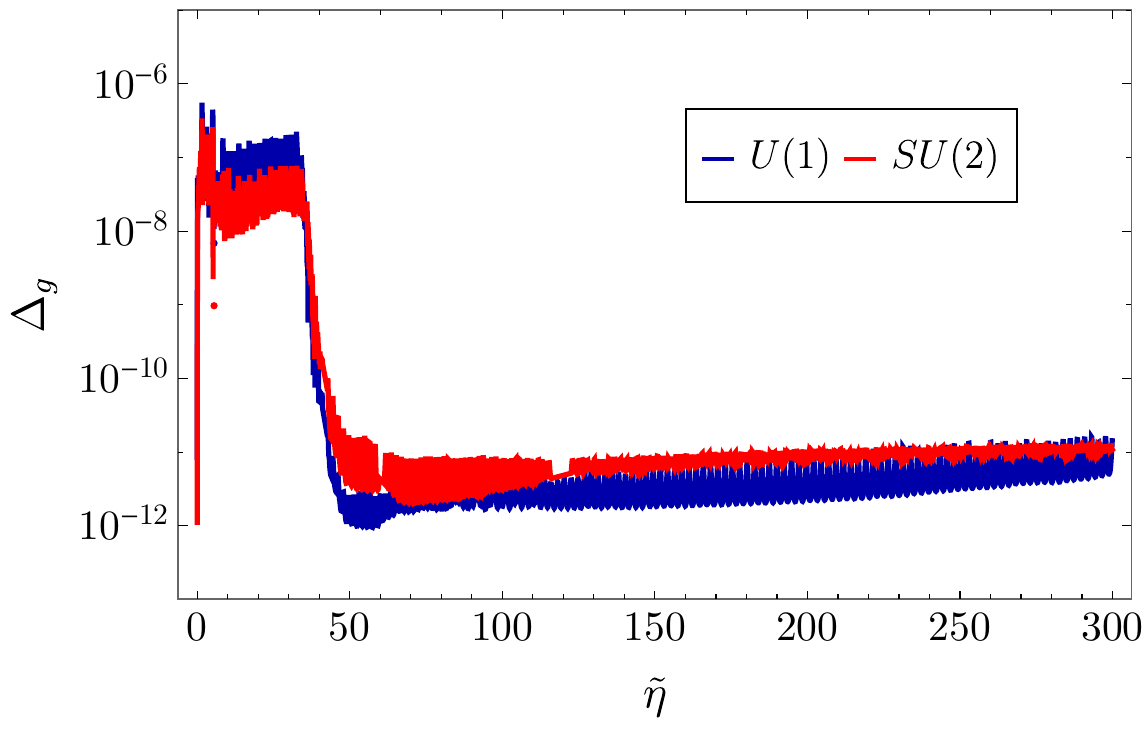}
    \caption{ Evolution of the relative error of the U(1) and SU(2) Gauss constraints for a simulation with gauge fields. We plot $\Delta _g \equiv {\left\langle\sqrt{(\mathrm{LHS}-\mathrm{RHS})^2}\right\rangle}/{\left\langle\sqrt{(\mathrm{LHS}+\mathrm{RHS})^2}\right\rangle}$, where $\mathrm{LHS}$ and $\mathrm{RHS}$ are the left and right hand sides of Eqs.~\eqref{eq:GaussU1-eom} [for U(1)] and \eqref{eq:GaussSU2-eom} [for SU(2)].
    The details of the simulated model are provided in Sect.~5.3 of the \CL user manual \cite{Figueroa:2021yhd}.} \label{fig:gausslaw}

\end{figure}

\CL can also simulate $SU(2)$ non-Abelian gauge fields coupled to charged doublet scalars. The $SU(2)$ doublet can be charged under $U(1)$, but for clarity, we discuss the non-Abelian sector in isolation. The relevant action is
\begin{eqnarray}
S = - \int d^4 x\sqrt{-g} \left(   (D_{\mu}\Phi )^{\dagger} (D^{\mu} \Phi) + \frac{1}{2}{\rm Tr}\{G_{\mu \nu}G^{\mu \nu}\} + V(\Phi)\right) \ ,
\label{eq:LagrangianII}
\end{eqnarray}
with $V(\Phi) \equiv V(|\Phi|)$ a potential term, where $\Phi$ is a $SU(2)$-doublet field
\begin{eqnarray} \label{eq:ChargedScalars}
 \Phi = \left(
\begin{array}{c}
\varphi^{(0)} \\ \varphi^{(1)} 
\end{array}
\right) =
{1\over\sqrt{2}}
\left(
\begin{array}{c}
\varphi_0 +i\varphi_1 \vspace*{0.1cm}\\ \varphi_2 +i\varphi_3 
\end{array}
\right) \,,
\end{eqnarray}
and where we have defined the covariant derivative and field strength tensor as
\bea
D_{\mu} &\equiv & 
\mathcal{I}\partial_\mu
- i g_B Q_B B_{\mu}^a \,T_a \ , \label{eq:CovDerivCont}\\
G_{\mu \nu} & \equiv & \partial_{\mu} B_{\nu} - \partial_{\nu} B_{\mu} - i[B_\mu,B_\nu] \ , \label{eq:GmnNonAb}
\eea
with $\mathcal{I}$ the $N\times N$ identity matrix, and $Q_B$ the non-Abelian charge of $\Phi$. Here $\{T_a\}$ are the generator of $SU(2)$.
\noindent The EOM of the system can be obtained from minimizing Eq.~(\ref{eq:LagrangianII}) as
\begin{eqnarray}
\Phi'' - a^{-2(1 - \alpha)} {\vv D}^{\,2}\Phi + (3 - \alpha)\frac{{a'}}{a}  {\Phi'} = - \frac{a^{2 \alpha} V_{,|\Phi|}}{2} \frac{\Phi}{|\Phi |} \ , \label{eq:higgsSU2-eom}
\\
(\mathcal{D}_0 )_{a b} (G_{0i})^b - a^{-2(1 - \alpha )} ( \mathcal{D}_j )_{a b} (G_{ji} )^b + (1 - \alpha) \frac{{a'}}{a} (G_{0i} )^b = a^{2 \alpha}(J_i)_a \ , \label{eq:SU2eom}
\\
(\mathcal{D}_i )_{a b} (G_{0i})^b = a^2(J_0)_a \,, \label{eq:GaussSU2-eom}
\end{eqnarray}
with the $SU(2)$ charged current defined as
\begin{eqnarray}
\label{eq:NonAbelianCurrent}
J_a^\mu \equiv 2g_BQ_B\mathcal{I}m [ \Phi^{\dag} T_a( D^{\mu} \Phi )]\,.
\end{eqnarray}
The Friedmann equations read now  \eqref{eq:sfEOM}
\begin{eqnarray}
\label{eq:FriedmannDDa}
\hspace*{-0.7cm}{a''\over a} = \frac{a^{2 \alpha}}{3 m_p^2}\big\langle (\alpha-2){K}_{\Phi} + \alpha {G}_{\Phi}) + (\alpha + 1)V+ (\alpha-1)({K}_{SU(2)} + {G}_{SU(2)}) \big\rangle \,,\\
\hspace*{-1.4cm} \left({a'\over a}\right)^2 =  \frac{a^{2 \alpha}}{3 m_p^2}\big\langle{K}_{\Phi} + {G}_{\Phi} + {K}_{SU(2)} + {G}_{SU(2)} + {V}\big\rangle \,, \label{eq:FriedNonAbEnc}
\end{eqnarray}
with the $SU(2)$ energy contributions defined as
\begin{eqnarray}
\label{eq:energy-contrib}
{K}_{\Phi} = \frac{1}{a^{2\alpha} } (D_0 \Phi )^\dag(D_0 \Phi) \ , \hspace*{1cm}
 {G}_{\Phi} = \frac{1}{a^2} \sum_i (D_i \Phi)^\dag(D_i \Phi) \ , \\
{K}_{SU(2)} = \frac{1}{2 a^{2 + 2 \alpha}}  \sum_{a,i} (G_{0i}^a)^2 \ , \hspace*{1cm}
{G}_{SU(2)} = \frac{1}{2 a^4}  \sum_{a,i,j<i}  (G_{ij}^a)^2  \,.
\end{eqnarray}
\CL can solve the EOM~(\ref{eq:higgsSU2-eom})-(\ref{eq:SU2eom}) together with Eq.~(\ref{eq:FriedmannDDa}) for the expansion of the Universe.
Here, \eqref{eq:GaussSU2-eom} is the Gauss constraint, which is preserved up to machine precision by the evolution algorithms implemented in \CLns, see Fig.~\ref{fig:gausslaw}. Eq.~\eqref{eq:FriedNonAbEnc} is used instead as a consistency check for ``energy conservation''.

The gauge field modules implemented in \CL can be used to simulate e.g.~the post-inflationary dynamics of the Standard Model, where the charged scalar represents the Higgs field \cite{Figueroa:2015rqa,Enqvist:2015sua,Figueroa:2016ojl}. It can also be used to study e.g.~preheating after hybrid inflation, where the symmetry breaking field is charged under a local gauge symmetry  \cite{Dufaux:2010cf}. An example of a \CL simulation including both Abelian and non-Abelian gauge fields is presented in Sect.~8 of \cite{Figueroa:2020rrl}.

\subsection{Gravitational waves} \label{subsec:GWs} 

Gravitational waves (GWs) are spatial perturbations of the background FLRW metric that are transverse and traceless. In cosmic time we write $ds^2 = - dt^2 + a^2 (t) ( \delta_{ij} + h_{ij}) {\rm d} x^i{\rm d} x^j$, so that $\partial_i h_{ij} = h_{ii} = 0$, and their dynamics are described by the EOM~\cite{Caprini:2018mtu}
\begin{eqnarray} 
\ddot{h}_{ij}  + 3 \left({\dot a\over a}\right)\dot{h}_{ij} - \frac{\nabla^2}{a^2} h_{ij} = \frac{2}{m_p^2 a^2} \Pi_{ij}^{\rm TT} \ , 
\end{eqnarray}
with $\Pi_{ij} \equiv T_{ij} - p  a^2(t) (\delta_{ij} + h_{ij})$ the anisotropic tensor of all matter fields sourcing GWs, which accounts for the deviation of the stress-energy tensor with respect to the perfect fluid form. Here the super-index ${\rm TT}$ denotes its transverse-traceless component.

Note that a {\rm TT}-projection is is a non-local operation in coordinate space and it is, therefore, time-expensive in a lattice. Conversely, in Fourier space one can easily define a local projection operator $\Lambda_{ij,kl} ({\bf \hat{k}}) $ such that $\Pi_{ij}^{\rm TT} (\vec{k}, t) = \Lambda_{ij,kl} ({\bf \hat{k}}) \Pi_{kl}  (\vec{k}, t)$ (see e.g.~Ref.~\cite{Figueroa:2011ye} for its explicit form). This allows to devise a procedure that solves the GW equations with a similar efficiency as for scalar or gauge field, originally proposed in  \cite{Garcia-Bellido:2007fiu}. In \CL we follow such procedure: we write the gravitational waves as $h_{ij} (\vec{k}, t) = \Lambda_{ij,kl} ({\bf \hat{k}}) u_{kl} (\vec{k}, t)$, and then solve the following equations in configuration space and in $\alpha$-time,
\begin{eqnarray}
\left\{\begin{aligned}
u_{i j}^{\prime} & =a^{\alpha-3}\left(\pi_{u}\right)_{i j} \ , \\
\left(\pi_{u}\right)_{i j}^{\prime} & =a^{1+\alpha} \nabla^2 u_{i j}+2 a^{1+\alpha} \Pi_{i j}^{\mathrm{eff}} \ , 
\end{aligned}\right.
\end{eqnarray}
where we have introduced an effective anisotropic tensor $\Pi_{ij}^{\rm eff}$ that only includes the parts of $\Pi_{\rm ij}$ with non-zero TT parts (see Eq.~\eqref{eq:anitensor} below). By solving this equation, we do not need to apply the TT-projection operator at each time step, but only when we need to compute $h_{ij}$, e.g.~to output the gravitational wave spectrum.

The energy density stored in a gravitational wave background is given by
\begin{eqnarray} 
\rho_{\rm GW} (t) \equiv \frac{m_p^2}{4 a^{2 \alpha}} \langle h'_{ij} (\vec{x}, t) h'_{i j} (\vec{x}, t) \rangle_ V \simeq \int \frac{d \rho_{\rm GW}}{d \log k} d \log k \ , \\
\frac{{\rm d} \rho_{\rm GW}}{{\rm d} \log k} \equiv \frac{m_p^2 k^3}{8 \pi^2 a^{2\alpha} V} \int \frac{{\rm d} \Omega_k}{4 \pi} h'_{ij} (\vec{k},t) h^{'*}_{ij} (\vec{k}, t) \ , 
\end{eqnarray}
where $\langle \dots \rangle_V$ is an average over the lattice volume $V$. \CL can compute the normalized energy density $\Omega_{\rm GW} \equiv \frac{1}{\rho_{\rm tot}} \frac{{\rm d} \rho_{\rm GW}}{{\rm d} \log k} $ at any time during the simulation, with $\rho_{\rm tot}$ the total energy density in the lattice. In the case of self-consistent expansion we have $\rho_{\rm tot} = \rho_c$, with $\rho_c$ the critical energy density of the system.

If the gravitational waves are sourced by all field species introduced above, $\{\phi$, $\varphi$, $\Phi$, $A_{\mu}$, $B_{\mu}^a \}$, the effective anisotropic tensor takes the form~\cite{PhDthesisFigueroa}
\begin{eqnarray}
    {\Pi}_{i j}^{\mathrm{eff}}={\nabla}_i {\phi} {\nabla}_j {\phi} & +2 \operatorname{Re}\left\{\left({D}_i {\varphi}\right)^*\left({D}_j {\varphi}\right)\right\}- (a^{-2 \alpha} {E}_i {E}_j+a^{-2} {B}_i {B}_j ) \label{eq:anitensor} \\
    & + 2 \operatorname{Re} \{ (D_i \Phi)^{\dagger} (D_j \Phi) \} - (a^{-2 \alpha}  E_i^a E_j^a + a^{-2} B_i^a B_j^a ) \ , \nonumber
\end{eqnarray}
where $E_i \equiv \partial_\eta A_i - \partial_i A_0$ and $B_i \equiv \epsilon_{ijk}\partial_jA_k$ are the electric and magnetic fields of the Abelian gauge sector, and $E_i^a$ and $B_i^a$ their respective non-Abelian counterparts (explicit expressions for $E_i^a$ and $B_i^a$ can be found e.g.~in~\cite{PhDthesisFigueroa}).

The possibility of simulating GWs sourced by scalar singlets was included in version 1.1 of \CLns, released in May 2022 (see \href{https://cosmolattice.net/assets/technical_notes/CosmoLattice_TechnicalNote_GWs.pdf}{\color{blue}\tt Technical Note 2}). The simulation of GWs sourced by a $U(1)$ gauge sector (formed by $\varphi$ and $A_{\mu}$) was implemented in version 1.2, released in June 2023 (see \href{https://cosmolattice.net/assets/technical_notes/CosmoLattice_TechnicalNote_GWsAbelianGauge.pdf}{\color{blue}\tt Technical Note 3}). We plan to implement the simulation of GWs from a $SU(2)$ sector (formed by $\Phi$ and $B_{\mu}^a$) in the near future.

\CL has been used to compute the gravitational waves produced by different early universe sources such as preheating and other post-inflationary resonance phenomena \cite{Cosme:2022htl,Cui:2023fbg,Mansfield:2023sqp}, cosmic strings \cite{Baeza-Ballesteros:2023say,Li:2023gil}, or oscillons \cite{Piani:2023aof}.

\subsection{Initial conditions}  \label{subsec:initcond}

To solve the EOM, we need to specify initial conditions for all fields. Denoting the initial time of a simulation as $\eta_*$, we split all fields in a spatially homogeneous mode and spatially varying fluctuations. To be explicit, e.g.~for a scalar field, we write 
\begin{eqnarray}
\phi ({\bf x}, \eta_* )  \equiv & \bar{\phi}_* + \delta \phi_* ({\bf x}) \ , \ \  \dot{\phi} ({\bf x}, \eta_* ) \equiv  \bar{\dot{\phi}}_* + \delta \dot{\phi}_* ({\bf x}) \label{eq:Scalar-init2}\ .
\end{eqnarray}
The value of the homogeneous modes depends on the specific model under investigation. For instance, in preheating scenarios, the inflation's homogeneous amplitude and velocity can  be computed at the end of slow-roll inflation. In all cases, the user must provide these values in a {\it parameter} input file, in accordance with the physics they want to investigate, see \CL manual~\cite{Figueroa:2021yhd}.

The initial spatial fluctuations of a scalar field are characterized by their spectrum, defined as the variance of the fluctuations via
\begin{eqnarray}\label{eq:SpectrumContinuum}
\left\langle \delta \phi^2 \right\rangle = \int d\log k~\Delta_{\delta \phi}(k)\,,~~~~~\Delta_{\delta \phi}(k) \equiv {k^3\over 2\pi^2}\mathcal{P}_{\delta \phi}(k)\,, \\ 
\left\langle {\delta \phi}_{\bf k}{\delta \phi}_{{\bf k}'} \right\rangle \equiv (2\pi)^3 \mathcal{P}_{\delta \phi} (k)\delta(\bf{k}-\bf{k}') \ .
\end{eqnarray}
\CL sets by default scalar fluctuations to mimic quantum vacuum fluctuations,
\begin{eqnarray}
    \mathcal{P}_{\delta \phi} (k) \equiv 
    \frac{1}{2 a^2\omega_{k,\phi}}\,,~~~~ \omega_{k,\phi} \equiv \sqrt{k^2 + a^2m_{\phi}^2} \,,~~~~ m_{\phi}^2 \equiv \frac{\partial^2 V}{\partial \phi^2}\Big|_{\phi = \bar{\phi}} \ . \label{eq:QuantumFlucts}
\end{eqnarray}
In practice, the fluctuations of scalar fields are set by writing the Fourier transform of a scalar field as
\bea
\delta  \phi ({  \bf k}) &=& \frac{1}{\sqrt{2}} (|\delta  \phi^{(l)} ({  \bf k})|  e^{i \theta^{(l)} ({   \bf {k}}) } + |\delta  \phi^{(r)} ({   \bf {k}})| e^{i \theta^{(r)} ({   \bf {k}}) }   ) \label{eq:fpr_influct} \ , \\
\delta  {\phi}' ({   \bf {k}})
&=& {1\over a^{1-
\alpha}}\left[\frac{i {\omega}_k}{\sqrt{2}}  \left(|\delta  \phi^{(l)} ({   \bf {k}})| e^{i \theta^{(l)} ({   \bf {k}}) } - |\delta  \phi^{(r)}  ({   \bf {k}})| e^{i \theta^{(r)} ({   \bf {k}}) }   \right)\right]  - \mathcal{H}  \delta  \phi ({   \bf {k}}) , \label{eq:fpr_influct2} \eea
where $|\delta  \phi^{(l,r)}|$ are random fields drawn from a Gaussian distribution with variance $\sigma_{\delta\phi}^2 =  \mathcal{P}_{\delta \phi} ({\bf k})$, with $\mathcal{P}_{\delta \phi}$ given in \eqref{eq:QuantumFlucts}. The variables $\theta^{(l)} ({\bf {k}})$ and $\theta^{(r)} ({\bf {k}})$ are two random independent phases that vary uniformly in the range $[0, 2\pi)$, from point to point in Fourier space~\cite{Figueroa:2020rrl}.

While in gauge theories the strategy is similar, the need to impose initial conditions that preserve the Gauss constraints~\eqref{eq:GaussU1-eom} and \eqref{eq:GaussSU2-eom}, makes it harder to implement in practice. The fluctuations need to be chosen in such a way that they respect the Gauss constraint(s), but there is no unique way to achieve this. In \CL we consider the following prescription for the gauge fields,
\bea
A_i ({\bf x}, \eta_* ) & = & 0 \ , \hspace{0.6cm} \dot{A}_i ({\bf x}, \eta_* )  \equiv  \delta \dot{A}_{i*} ({\bf x}) \ ,  \label{eq:Inflc1}\\
B_i^a ({\bf x}, \eta_* ) & = & 0 \ , \hspace{0.6cm} \dot{B}_i^a ({\bf x}, \eta_* )  \equiv  \delta \dot{B}_{i*}^a ({\bf x}) \ ,  \label{eq:Inflc2}
\eea
generating initial electric field fluctuations and vanishing magnetic field ones. In particular, once the fluctuations in the scalar fields have been set, we solve for the initial fluctuations of the electric fields from the initial Gauss laws \eqref{eq:GaussU1-eom} and \eqref{eq:GaussSU2-eom} in momentum space (we set the initial scale factor to one for simplicity)~\cite{Figueroa:2020rrl}
\bea {k}^i {A}'_i ({\bf k}) = J_0^A ({\bf k}) \ , \hspace{0.8cm} {k}^i {B}_i^{a'} ({\bf k}) = J_0^a ({\bf k}) \ . \label{eq:kAi1} \eea 

A constraint that needs to be imposed to set scalar field fluctuations properly is that either Abelian and non-Abelian total charges must vanish. For clarity, we show this only for the $U(1)$ case (the procedure is easily generalized to $SU(2)$).  We want a vanishing initial charge $J_0^A ({\bf k} ={\bf 0} ) = \int d^3 {\bf x}\, J_0^A ({\bf x}) = 0$, which implies \cite{Figueroa:2020rrl}
\begin{eqnarray}
\label{eq:totalchargezero}
 \hspace{-0.5cm}\int d^3 {\bf k}\, \mathcal{R}e [\varphi_0^* ({\bf k}) \varphi'_{1} ({\bf k}) - \varphi'_0 ({\bf k}) \varphi_1^* ({\bf k}) + \varphi_2^* ({\bf k}) \varphi'_3 ({\bf k}) - \varphi'_2 ({\bf k}) \varphi_3^* ({\bf k}) ] = 0 \ .  
\end{eqnarray}
The fluctuations of the scalar components $\varphi_n$ are then set in a similar fashion as for scalar singlets. We start by separating the homogeneous mode and fluctuations as
\bea
 \varphi_n ({\bf x}, t_* ) & \equiv& \frac{|\varphi_*|}{\sqrt{2}}  +  \delta \varphi_{n*} ({ \bf x}) \ , \label{eq:CPhi_init1}\\
  \dot{\varphi}_n ({\bf x}, t_* ) &\equiv & \frac{|\dot{\varphi}_*|}{\sqrt{2}}  +  \delta \dot{\varphi}_{n*} ({ \bf x}) \label{eq:CPhi_init2} \ ,
\eea
and we impose the following functional form for the fluctuations,
\begin{align}
\delta {\varphi}_n({  \bf k}) &= \frac{1}{\sqrt{2}} \left(|\delta {\varphi}_{n}^{(l)} ({  \bf k})|  e^{i \theta_{n}^{(l)} ({  \bf k}) } + |\delta {\varphi}_{n}^{(r)} ({  \bf k})| e^{i \theta_{n}^{(r)} ({  \bf  k}) }   \right) \label{eq:fpr_influct3} \ , \\
\delta {\varphi}'_n ({  \bf k}) &= a^{1-\alpha}\left[\frac{1}{\sqrt{2}} i {\omega}_{k,n} \left(|\delta {\varphi}_{n}^{(l)} ({  \bf  k})| e^{i \theta_{n}^{(l)} ({  \bf  k}) } - |\delta {\varphi}_{n}^{(r)}  ({  \bf  k})| e^{i \theta_{n}^{(r)} ({  \bf  k}) }  \right)\right]  - {\mathcal{H}} \delta {\varphi}_{n} ({  \bf k})\ . \label{eq:fpr_influct4}
\end{align}
Similarly as for singlet scalars, we introduce $|\delta {\varphi}_{n}^{(r)}  ({  \bf  k})|$ as random Gaussian fields with variance set by the power spectrum of quantum vacuum fluctuations~(\ref{eq:QuantumFlucts}). The total charge constraint $\eqref{eq:totalchargezero}$ can then be satisfied by choosing $\delta \varphi_{0}^{(l)} ({\bf k})= \delta \varphi_{0}^{(r)} ({\bf k})$,  $\delta \varphi_{1}^{(l)} ({\bf k})= \delta \varphi_{1}^{(r)} ({\bf k})$, and $\theta_{1}^{(r)} ({\bf k})= \theta_{0}^{(r)} ({\bf k})+ \theta_{1}^{(l)}({\bf k})  - \theta_{0}^{(l)} ({\bf k})$.

We note that this choice of initial conditions is not unique and in a coming update to \CLns, the user will be able to specify their own initial power spectra by means of external text files, see Sect.~\ref{sec:arbitraryspec}.

\subsection{Output} \label{sec:output}

\begin{figure}
    \centering
    \includegraphics[width=.48\textwidth]{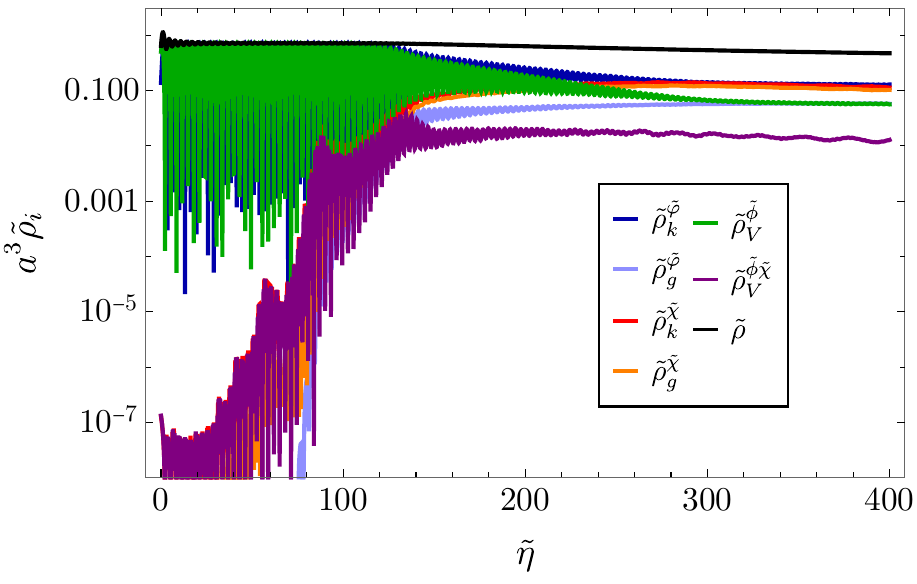}\,\,
    \includegraphics[width=.48\textwidth]{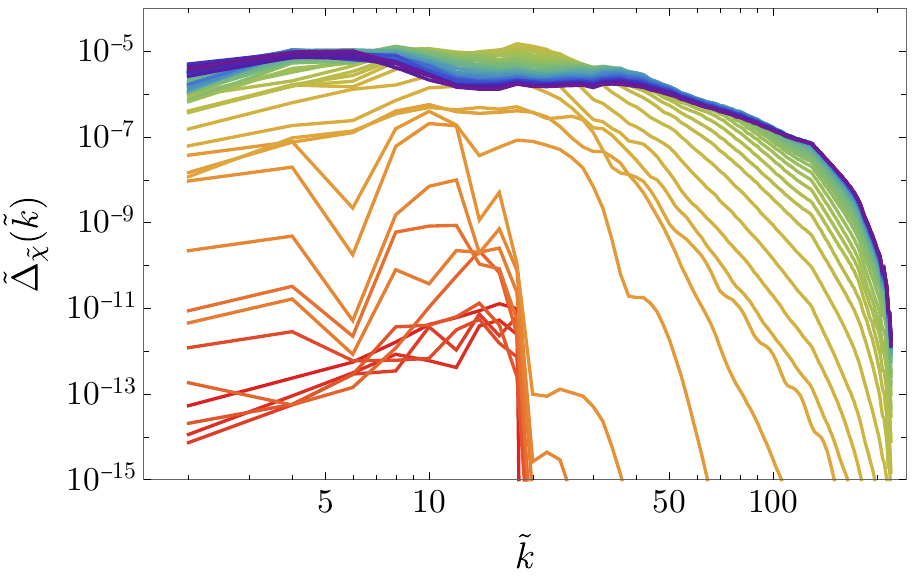} \vspace*{0.2cm}\\
      \centering
       \,\,\,\,\,\,\,\,\,\,\,\,\,\,\,   \includegraphics[width=.43\textwidth]{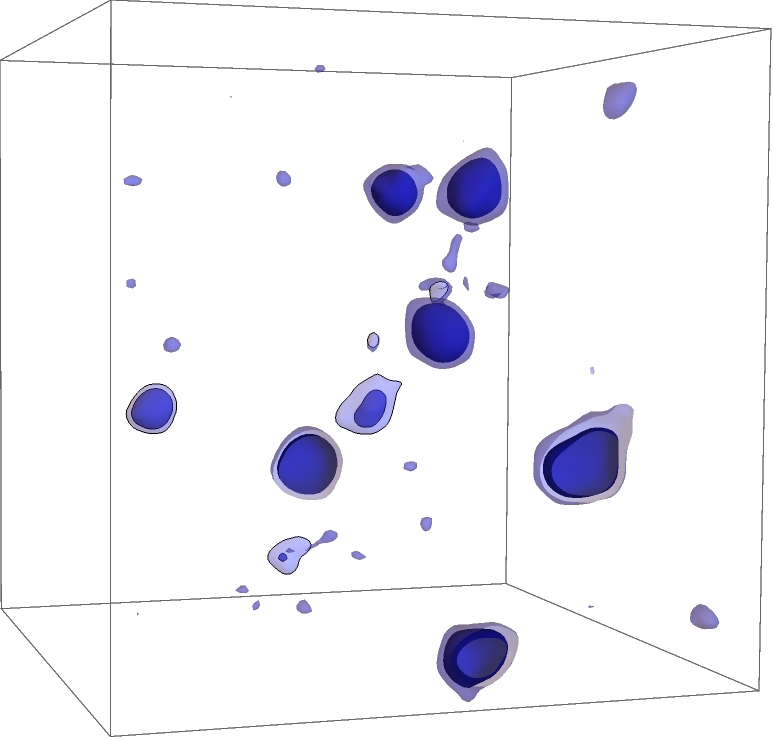}\,\,\,\,\,\,\,
    \includegraphics[width=.45\textwidth]{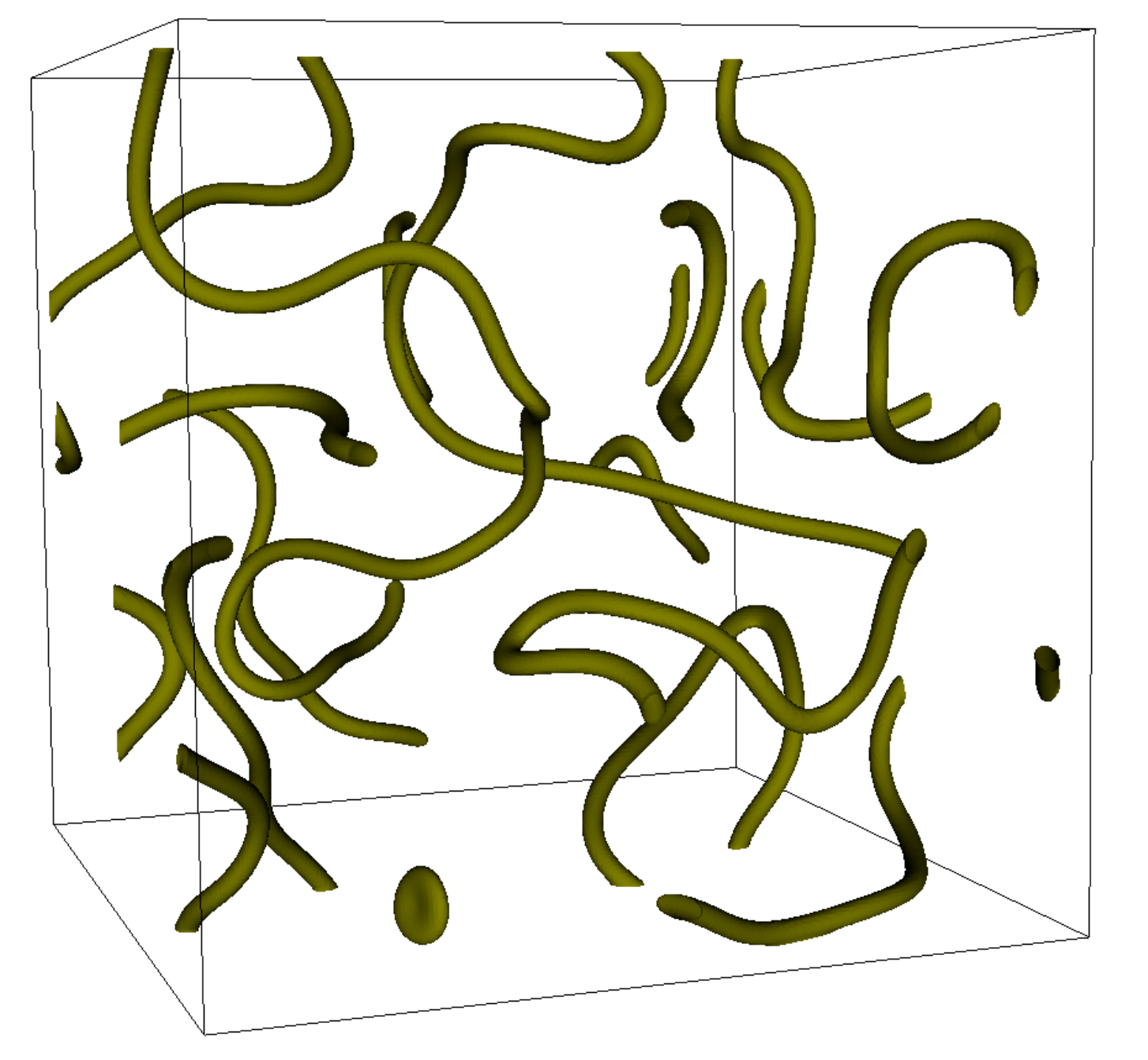}     \label{fig:CLoutput}
    \caption{ Examples of different kinds of outputs generated by \CLns. Top-left: Volume average of the energy density and its components (kinetic, gradient and potential) during preheating, see \cite{Antusch:2021aiw}. Top-right: Evolution of the daughter field power spectrum in the same set-up. Bottom-left: Snapshot of the oscillons generated after hilltop inflation, see \cite{Antusch:2019qrr}. Bottom-right: Snapshots of a network of cosmic strings, see \cite{Baeza-Ballesteros:2023say}.}

\end{figure}

\CL can generate three different kinds of output at any time during the evolution:

\begin{itemize}
    \item \textit{Volume averages:} Average of a given quantity over the lattice volume. For example, \CL outputs the volume average of the field amplitudes and energy density components by default.
    \item \textit{Spectra:} Values of the binarized power spectrum of a given field quantity  $\Delta_{f}(k)$ as a function of $k$. For example, in the case of scalar fields, \CL outputs $\Delta_{\delta \phi}(k)$, $\Delta_{\delta \phi'}(k)$, and the occupation number by default. The user can freely choose the width of the bins and the output format (either text or HDF5).
    \item \textit{Snapshots:} 3-dimensional distributions of a given field quantity (such as the energy density components) in HDF5 format.
\end{itemize}
The \CL user manual provides a comprehensive list of the outputs generated by the code for the different matter sectors \cite{Figueroa:2021yhd}. The implementation of new outputs is, however, a straightforward process.  Note that the code allows to specify different printing frequencies for the three kinds of output, which is convenient because e.g.~the computation of field spectra is more time expensive than volume averages. Fig.~\ref{fig:CLoutput} shows examples of different kinds of output obtained with \CLns.

\section{Future update of \CLns, Part I: new physics} \label{sec:futurep1}

\subsection{Axions}
\label{subsec:Axions}

In beyond the Standard Model (BSM) scenarios, axion-like particles (ALPs) appear as pseudo Nambu-Goldstone boson fields of spontaneously broken global symmetries. Originally proposed as a solution to the strong CP problem~\cite{Peccei:1977hh, Peccei:1977ur, Weinberg:1977ma, Wilczek:1977pj}, ALPs are also invoked in cosmology as, e.g.~dark matter (DM)~\cite{Abbott:1982af, Dine:1982ah,Preskill:1982cy,Hui:2016ltb} or inflaton candidates~\cite{Freese:1990rb,Pajer:2013fsa, Adshead:2015pva, Domcke:2019qmm, Adshead:2019lbr, Figueroa:2023oxc}. Axions also appear in String theory as generic particles in the low-energy spectrum~\cite{Witten:1984dg,Svrcek:2006yi,Arvanitaki:2009fg,Marsh:2015xka}, with some string constructions leading to an axion monodromy~\cite{McAllister:2008hb,Silverstein:2008sg,Marchesano:2014mla,Blumenhagen:2014gta,Hebecker:2014eua,McAllister:2014mpa}, where an ALP can probe multiple non-degenerate minima of its potential. This has been exploited in the context of inflation~\cite{McAllister:2008hb,Silverstein:2008sg,Marchesano:2014mla}, and as a dynamical solution to the electroweak (EW) hierarchy problem~\cite{Graham:2015cka}. In summary, axions are common ingredients in theoretical constructions of inflation and BSM physics.

Due to shift symmetry, the interaction of an ALP with other species is very restricted. For example, the lowest dimensional interaction between an ALP $\phi$ and an Abelian gauge sector is 
\begin{eqnarray}\label{eq:phiFtildeF}
\mathcal{L}_{\rm int} \propto \phi F_{\mu \nu} \tilde F^{\mu \nu}\,,
\end{eqnarray}
with $F_{\mu \nu}$ the field strength of the gauge field $A_\mu$, and $\tilde F_{\mu \nu}$ its dual. ALPs can excite very strongly gauge field quanta through such Chern-Simons coupling. For example, in {\it axion inflation} scenarios, where 
the ALP is identified with the inflaton, the gauge field can be highly excited during inflation, leading to a rich phenomenology that includes: the production of a sizeable background of chiral GWs~\cite{Sorbo:2011rz, Barnaby:2011qe, Cook:2013xea,Adshead:2013qp,Bastero-Gil:2022fme,Garcia-Bellido:2023ser}, which can be searched for with direct detection GW experiments~\cite{Cook:2011hg,Anber:2012du,Domcke:2016bkh,Bartolo:2016ami}; the creation of large scalar perturbations~\cite{Barnaby:2010vf,Barnaby:2011qe,Barnaby:2011vw,Cook:2011hg, Barnaby:2011qe,Pajer:2013fsa,Domcke:2020zez,Caravano:2022epk}, which can be probed by the cosmic microwave background (CMB)~\cite{Barnaby:2010vf,Meerburg:2012id,Sorbo:2011rz} and searches for primordial black holes (PBHs)~\cite{Linde:2012bt, Pajer:2013fsa, Bugaev:2013fya, Cheng:2015oqa, Garcia-Bellido:2016dkw,Garcia-Bellido:2017aan,Domcke:2017fix,Cheng:2018yyr,Ozsoy:2023ryl}; efficient magnetogenesis~\cite{Garretson:1992vt, Anber:2006xt, Adshead:2016iae, Durrer:2023rhc}, baryon asymmetry~\cite{Giovannini:1997eg,Anber:2015yca,Fujita:2016igl,Kamada:2016eeb,Jimenez:2017cdr,Cado:2022evn} and reheating~\cite{Adshead:2015pva,Cuissa:2018oiw} mechanisms, which can also be naturally realized in these scenarios. Furthermore, thanks to interaction~(\ref{eq:phiFtildeF}), a dark matter ALP can also excite very efficiently a gauge field during the post-inflationary evolution of the universe~\cite{Domcke:2019qmm,Machado:2018nqk,Machado:2019xuc,Ratzinger:2020oct}, possibly also resulting in the production of GWs~\cite{Banerjee:2021oeu,Madge:2021abk}.

Given the richness of the above phenomenology, and the interest to consider new axion applications, we plan to release an {\it axion} module for \CLns, which will be dedicated to simulate the dynamics expected from an action ${\cal S}_{\rm ax} = -\int {\rm d}x^4 \sqrt{-g}\big\lbrace \frac{1}{2}\partial_\mu \phi\partial^\mu\phi+V(\phi) + \frac{1}{4}F_{\mu\nu}F^{\mu\nu} + \frac{\alpha_{\Lambda}}{4}\frac{\phi}{m_p} F_{\mu\nu}\tilde F^{\mu\nu} \big\rbrace$, with  $\alpha_{\Lambda}$ the axion-gauge coupling. The variation of $S_{\rm tot} = S_{\rm g} + S_{\rm m}$ with $S_{\rm g} \equiv \int {\rm d}x^4 \sqrt{-g}\,\frac{1}{2}m_p^2 R$ standard Hilbert-Einstein gravity, leads to a system of equations of motion which, considering a flat expanding background, read as  (here $t$ represents cosmic time, $\dot{} \equiv d/d t$ and $H(t)={\dot a/a}$)
\begin{eqnarray}\label{eq:AxionInfEOM}
\left.
\begin{array}{rcl}
\ddot\phi &=& -3H\dot\phi+\frac{1}{a^2}\vec\nabla^2\phi-V_{,\phi}+\frac{\alpha_\Lambda}{a^3 m_p}\vec{E}\cdot\vec{B}\,,\label{eqn:eom1}\vspace*{1mm}\\
\dot{\vec{E}} &=& -H\vec{E}-\frac{1}{a^2}\vec{\nabla}\times\vec{B}-\frac{\alpha_\Lambda}{a m_p}\Big(\dot\phi\vec{B}-\vec{\nabla}\phi\times\vec{E}\Big),\label{eqn:eom2}\vspace*{2mm}\\
\ddot{a} &=&
-\frac{a}{3m_p^2}\big\langle 2\rho_{\rm K}-\rho_{\rm V}+\rho_{\rm EM} \big\rangle\,, 
\label{eqn:eom3}\vspace*{2mm}\\
\vec{\nabla}\cdot\vec{E} \,\,&=& -\frac{\alpha_{\Lambda}}{a m_p}\vec{\nabla}\phi\cdot\vec{B}\,,\hspace*{3cm}{\rm\tt [Gauss~law]}\label{eq:Gauss}\vspace*{2mm}\\
H^2\, &=& ~\frac{1}{3m_p^2}\big\langle \rho_{\rm K}+\rho_{\rm G}+\rho_{\rm V}+\rho_{\rm EM}\big\rangle\,,{\rm\tt ~~[Hubble~law]}
\label{eq:Hubble}
\end{array}\right\rbrace
\end{eqnarray}
with $\vec B \equiv \vec\nabla \times \vec A$ the magnetic field, $\vec E \equiv \partial_t{\vec A}$ the electric field (in the temporal gauge $A_0 = 0$), and where the electromagnetic and inflaton's kinetic, potential and gradient energy densities are given by $\rho_{\rm EM} \equiv \frac{1}{2a^4}\langle a^2\vec E^2+\vec B^2\rangle$, $\rho_{\rm K} \equiv \frac{1}{2}\langle\dot{\phi}^2\rangle$, $\rho_{\rm V} \equiv \langle V \rangle$, and $\rho_{\rm G} \equiv \frac{1}{2a^2}\langle(\vec\nabla\phi)^2\rangle$, respectively, with $\langle ... \rangle$ denoting volume averaging. While the first three equations describe the system dynamics, the last two represent constraint equations.

Though we have presented here the continuum dynamics in cosmic time $t$ for easiness of the reader, the new module will operate in any $\alpha$-time $\eta$, including e-folding $N = \log a(t)$. A working implementation of these dynamics is actually ready and has been successfully tested and used in Ref.~\cite{Figueroa:2023oxc}. 

\subsection{Non-minimal gravity}
\label{subsec:nonminimal-gravity}

In {\it gravitational (p)reheating} scenarios, a spectator scalar field $\chi$ is often non-minimally coupled to gravity, via an interaction of the form $\xi \chi^2 R$, with $R$ the Ricci scalar and $\xi$ a dimensionless coupling constant. The presence of such coupling is actually required by the renormalizability of scalar fields in a curved spacetime~\cite{Parker2009,Birrell1984}. For example, early formulations of gravitational reheating~\cite{Ford:1986sy,Spokoiny:1993kt,Damour:1995pd} included two ingredients, the excitation of a non-minimally coupled scalar field, $\chi$, towards the end of inflation, and the occurrence of a period of {\it kination} domination (KD) following the end of inflation. These are the basic ingredients of e.g.~{\it Quintessential inflation} scenarios~\cite{Peebles:1998qn,Peloso:1999dm,Huey:2001ae,Majumdar:2001mm,Dimopoulos:2001ix,Wetterich:2013jsa,Wetterich:2014gaa,Hossain:2014xha,Rubio:2017gty}, where the initially subdominant energy of $\chi$, assumed in the form of radiation, eventually becomes the dominant energy component of the Universe. While the original idea has been shown to be problematic~\cite{Figueroa:2018twl}, a variation known as {\it Ricci reheating}~\cite{Figueroa:2016dsc,Opferkuch:2019zbd,Dimopoulos:2018wfg,Bettoni:2021zhq} corrects naturally the problems, thanks to realizing that the non-minimally coupled scalars are exponentially excited during KD, rather than behaving as radiation, as originally assumed in~\cite{Ford:1986sy,Spokoiny:1993kt,Damour:1995pd}.

Furthermore, 
non-minimally coupled daughter scalars can also be naturally excited after inflation, if the inflaton potential is characterized by a monomial shape. In this case, the oscillatory behavior of $R$ leads to a tachyonic excitation of non-minimally coupled scalars, whenever $R$ becomes negative within each oscillation. This mechanism, introduced and coined as {\it geometric preheating} in Ref.~\cite{Bassett:1997az}, has been later considered in Ref.~\cite{Tsujikawa:1999jh}, in higher order curvature inflationary models~\cite{Tsujikawa:1999iv,Fu:2019qqe}, in multi-field inflationary scenarios~\cite{DeCross:2015uza,DeCross:2016fdz,DeCross:2016cbs,Nguyen:2019kbm,vandeVis:2020qcp}, and in dark matter production~\cite{Lebedev:2022vwf,Garcia:2023qab}. 

Most studies typically work
out the dynamics of non-minimally coupled scalar fields in the Einstein frame, where gravity is simply described by a Hilbert-Einstein term, and analytic calculations can be employed more easily. It is not well understood, however, to what extent the Jordan and the Einstein frames are equivalent at the full quantum level, as the conformal factor to change from the Jordan to the Einstein frame is a local function of the non-minimally coupled field, which is often a quantum field. 
To avoid any ambiguity in this respect, Ref.~\cite{Figueroa:2021iwm} has recently proposed a new technique that can solve the dynamics in the Jordan frame of an non-minimally coupled scalar field, while considering an expanding background sourced by all fields present. This includes situations when the dynamics become fully inhomogeneous and/or fully non-linear due to backreaction of the non-minimally coupled species. 

Given the variety of interesting applications of non-minimal gravitationally coupled fields, we plan to release a {\it non-minimal gravity} module for \CLns, dedicated to simulate the dynamics of non-minimally coupled (NMC) scalar species in the Jordan frame, including the case of self-consistent evolution of the expanding background as sourced by the non-minimally coupled species. Following Ref.~\cite{Figueroa:2021iwm}, we consider an action $\mathcal{S}_{\rm tot} = \mathcal{S}_{\rm NMC} + S_{\rm g} + S_{\rm m}$, where $\mathcal{S}_{\rm NMC} = 
-\int d^{4}x \sqrt{-g} \left(\frac{1}{2}\xi R \phi^{2} +\frac{1}{2} g^{\mu\nu}\partial_{\mu}\phi\partial_{\nu}\phi + V(\phi,{\{\varphi_{\rm m}\}})\right)$ describes the NMC scalar dynamics, $S_{\rm g} \equiv \int {\rm d}x^4 \sqrt{-g}\,\frac{1}{2}m_p^2 R$ is the standard Hilbert-Einstein term, and $S_{\rm m} \equiv \int {\rm d}x^4 \sqrt{-g}\,\mathcal{L}_{\rm m}({\{\varphi_{\rm m}\}})$ represents minimally coupled sectors. If the latter consists e.g.~of minimally coupled scalars $\{\varphi_{\rm m}\}$, the variation of $\mathcal{S}_{\rm tot}$ leads to a set of equations of motion which, in a flat expanding background and written in generic $\alpha$-time, read 
\begin{align}
{\tt [Minimally~coupled]} & \left\lbrace
\begin{array}{l}
\varphi_m' = a^{\alpha-3}\pi_{\varphi_m}\,, \vspace*{0.2cm}\\ 
\pi'_{\varphi_m} =  a^{1+\alpha}\, {\nabla}^2 \varphi -a^{3+\alpha} V_{,\varphi_m}\,,
\end{array}\right.
\label{eq:EOMpi}\\
{\tt [Non-minimally~coupled]} & \left\lbrace
\begin{array}{l}
\phi' = a^{\alpha-3}\pi_{\phi}\,, \vspace*{0.2cm}\\ 
\pi'_{\phi} =  a^{1+\alpha}\, {\nabla}^2 \phi -a^{3+\alpha} \left(\xi  R \phi + V_{,\phi}  \right)\,,
\end{array}\right.
\label{eq:EOMpiNMC}\\
{\tt [Expanding~background]} & \left\lbrace
\begin{array}{l}
a' = a^{\alpha-1}\pi_{a}\,, \vspace*{0.2cm}\\  
\pi_a' = \frac{a^{{2+\alpha}}}{6}   R\,,
\end{array}
\right.
\label{eq:pia}
\end{align}
with
\begin{align}
    R &=\frac{1}{m_p^2} \left[\frac{2\left(1-6\xi \right) \big(  E_G^{\phi} - E_K^{\phi}\big)  + 4\langle  V\rangle- 6\xi\langle \phi \, V_{,\phi}\rangle+(\bar{\rho}_{\rm  m}-3\bar{p}_{\rm  m})}{1 + \left(6\xi -1\right)\xi \langle\phi^2\rangle/m_p^2}\right]\label{eq:Rnew} \,,
\end{align}
and where volume-averaged kinetic and gradient energy densities of the NMC field are given by $E_K^{\phi} = \frac{1}{2a^6}\langle \pi_{\phi}^2 \rangle$ and $E_G^{\phi} = \frac{1}{2a^2} \sum_{i} \langle \partial_{i}\phi\partial_{i}\phi  \rangle$. 

We note that in Eq.~(\ref{eq:Rnew}) we have used the {\it trace} of the energy momentum of the minimally coupled matter fields, $\langle T_{\rm m} \rangle =3\bar p_{\rm  m} - \bar\rho_{\rm  m}$, without specifying the nature of the minimally coupled sector. Even though, for clarity, we wrote down the EOM above considering minimally coupled singlet scalars, c.f.~Eq.~(\ref{eq:EOMpi}), we could have also considered a gauge theory. In that case we just need to substitute Eq.~(\ref{eq:EOMpi}) by the EOM describing the dynamics of charged scalars and gauge fields, see sections~\ref{subsec:Abelian}, \ref{subsec:NonAbelian}. The algorithm would still read the same, with the piece $(3\bar p_{\rm  m} - \bar\rho_{\rm  m})$ in Eq.~(\ref{eq:Rnew}) receiving now contributions from all minimally coupled scalar and gauge fields. 
Furthermore, generalization to multiple non-minimally coupled scalars can be obtained straight forwardly by summing over the terms with non-minimal coupling in \eqref{eq:Rnew}.

A working implementation of the EOM of non-minimally coupled scalars is actually ready and has been successfully tested and used in Ref.~\cite{Figueroa:2021iwm,Figueroa:2024asq}.

\subsection{Cosmic strings and other defects}
\label{subsec:cosmicDefects}

Cosmic defects are stable energy configurations that may form in the universe, whenever some scalar field(s) acquire, upon spontaneous symmetry breaking, a non-zero expectation value within a topologically ``non-trivial'' vacuum manifold~\cite{Kibble:1976sj}. There can be global or local defects, depending on whether the symmetry broken is global or gauge. Defect networks are expected to reach a {\it scaling} regime, characterised by the mean separation of defects tracking the cosmological horizon. 

One of the most interesting defect cases is that of cosmic strings, which are naturally predicted in a variety of field theory and superstring early Universe scenarios~\cite{Kibble:1976sj,Kibble:1980mv,Vilenkin:1984ib,Hindmarsh:1994re,Jeannerot:2003qv,Copeland:2009ga,Copeland:2011dx,Vachaspati:2015cma}. A network of cosmic strings consists primarily of `long' (infinite) strings stretching across the observable universe, and loops of string. In the traditional picture the long string density decreases as they intercommute forming loops, which in turn decay mainly into gravitational waves (GWs). The combined incoherent emission of GWs from many loops leads to a GW background (GWB)~\cite{Vilenkin:1981bx,Hogan:1984is,Vachaspati:1984gt}. However, it is worth stressing that such view is based primarily on the Nambu-Goto (NG) approximation, which considers the strings as effectively infinitely thin line-like objects. Simulations of field theory strings~\cite{Matsunami:2019fss,Saurabh:2020pqe,Hindmarsh:2021mnl,Baeza-Ballesteros:2023say} indicate that strings decay into particles rather fast, and while the particle decay channel dominates clearly over the GW channel in the case of global string loops~\cite{Baeza-Ballesteros:2023say}, the question is not settled for local string loops~\cite{Matsunami:2019fss,Hindmarsh:2021mnl}.

Independently of the type of cosmic defects and of their origin, GWs are always emitted as the network's  energy-momentum tensor adapts itself to maintaining scaling~\cite{Figueroa:2012kw,Figueroa:2020lvo}. In the case of cosmic strings, on top of the GWs emitted from the long strings during the scaling dynamics, there is also a GW emission from subhorizon loop dynamics. For global strings, the GW signal is expected to be very weak if scaling is exact~\cite{Figueroa:2012kw,Figueroa:2020lvo}, whereas the amplitude is enhanced at cosmological scales~\cite{Chang:2019mza,Gouttenoire:2019kij,Gorghetto:2021fsn,Chang:2021afa,Servant:2023mwt} if log-violations of scaling~\cite{Gorghetto:2018myk,Gorghetto:2020qws,Buschmann:2021sdq} are present\footnote{In the case of axion/global string networks, logarithmic corrections to scaling have been claimed on the basis of fits to recent global string simulations~\cite{Gorghetto:2018myk,Gorghetto:2020qws,Buschmann:2021sdq,Buschmann:2021sdq}. Refs.~\cite{Hindmarsh:2019csc,Hindmarsh:2021vih,Hindmarsh:2021zkt} argue, however, that the apparent deviation of scaling corresponds simply to an early transient stage towards scaling.}. In the case of local strings the network GW emission is expected to be very subdominant compared to the GW emission from loops in the Nambu-Goto picture~\cite{Vilenkin:1981bx,Vachaspati:1984gt,Auclair:2019wcv}, whereas such result might be challenged in the case of field theory local string networks, depending on the loop configuration~\cite{Matsunami:2019fss,Hindmarsh:2021mnl}. 

Given the great interest in settling the correct details about scaling and GW emission from cosmic string networks, we plan to release a {\it cosmic defect} module for \CLns, dedicated to simulate the dynamics of cosmic defect networks. For example, for a global string network we will follow the recipe proposed in Ref.~\cite{Hindmarsh:2021vih}, where initial conditions start from realizations of a random Gaussian complex field $\phi = (\phi_1+i\phi_2)/\sqrt{2}$, with power spectrum
\begin{eqnarray}\label{eq:initialPS}
\mathcal{P}_{\phi_i}(k)=\frac{k^3 v^2 \ell_\text{str}^3}{\sqrt{2\pi}}\exp\left(-\frac{1}{2}k^2\ell_\text{str}^2\right)\,, \hspace{0.5cm} i = 1,2
\end{eqnarray}
normalized so that $\langle \phi_1^2 \rangle + \langle \phi_2^2\rangle =v^2$ is the vacuum expectation value ($vev$). Here $\ell_\text{str}$ is a correlation length that controls the string density of the resulting network. As the field configuration obtained like this is initially too energetic, to damp the excess energy it is customary to evolve the initial configuration with a {\it diffusion} process. Assuming a Mexican hat potential $V(\phi) = \lambda\left(|\phi|^2-v^2/2\right)^2$ with $\lambda$ the self-coupling of $\phi$, one can run a diffuse equation as
\begin{eqnarray}\label{eq:diff}
\Gamma_{\rm D}\,\phi_i'-\nabla^2\phi_i=-\lambda\left(\phi_1^2+\phi_2^2-v^2\right)\phi_i\,, \hspace*{1cm} i = 1,2\,,
\end{eqnarray}
where $\Gamma_{\rm D}$ is a diffusion rate, typically fixed to the characteristic scale of the problem $\Gamma_{\rm D} = \sqrt{\lambda}v$ for convenience. With that choice, diffusing the field configuration during a time scale $\mathcal{O}(10)$ times larger than the time it takes for a ray of light to go through the string core, is typically enough to leave a smooth string configuration. 

After diffusion, the string network could be evolved directly in an expanding background, say in RD. However, this is not convenient because the physical string width is constant, and hence its comoving core width shrinks as $\propto 1/a$. This implies that as expansion goes on, we will reach a moment when we no longer have enough resolution to resolve the string cores. To prevent this problem, one solution is to introduce a new phase after diffusion, during which the comoving string width is forced to increase in time. This corresponds to a string-core resolution-preserving approach~\cite{Press:1989yh}, during which the coupling $\lambda$ is promoted to a time dependent parameter, $\lambda=\lambda_0a^{2(s-1)}$. The comoving string width evolves then as $w=w_0 a^{-s}$, with tunable parameter $s$. In this case, the EOM of the field read
\begin{eqnarray}
\label{eq:EoMCSfat}
\phi_i''+2\frac{a'}{a}\phi_i'-\nabla^2\phi_i =-a^{2(s+1)}\lambda\left(\phi_1^2+\phi_2^2-v^2\right)\phi_i\,, ~~~~i = 1,2\,.
\end{eqnarray}

In practice, after diffusion, we evolve Eq.~(\ref{eq:EoMCSfat}) e.g.~in RD with $a=\tau/\tau_0$, with $\tau_0$ the moment when such evolution begun. Such background expansion is maintained for a half-light-crossing time of the lattice, i.e.~$\Delta \tau_\text{HLC}=L/2$, with $L$ the comoving length of the lattice. To avoid losing resolution of the string cores, we consider first a {\it fattening} period with $s = -1$ ($\lambda\rightarrow a^4\lambda$) during a time interval $\Delta \tau_\text{fat}$. At $\tau \geq \tau_0 + \Delta \tau_\text{fat}$, we switch to {\it physical} evolution with $s = 0$ ($\lambda \rightarrow const.$). By choosing $\Delta \tau_\text{fat}=\sqrt{\tau_0(\tau_0+\Delta \tau_\text{HLC})}$, we guarantee that the system arrives at $\tau_0+\Delta \tau_\text{HLC}$ in a configuration such that the comoving string-core width is equal to the initial physical width at the onset of background evolution (end of diffusion). Typically, during physical evolution within the period $\tau_0 + \Delta \tau_\text{fat} \leq \tau \leq \tau_0 + \Delta \tau_\text{HLC}$,  networks approach the scaling regime, with the mean string separation growing almost linearly in conformal time and the mean square velocity resulting constant.

A working implementation of the above procedure for global strings is actually ready and has been successfully tested and used in Ref.~\cite{Baeza-Ballesteros:2023say}. We plan to make a {\it cosmic defect} module available as part of \CLns, 
including scaling algorithms for global and local strings, and possibly for other defects as well, such as e.g.~domain walls. 

\subsection{Magneto hydrodynamics (MHD)}

First-order phase transitions (1stO-PhTs) proceed through bubble nucleation, which then grow and merge~\cite{Witten:1980ez,Guth:1981uk,Steinhardt:1981ct}. The collision of the bubbles is a violent
process that can lead to sound (pressure) waves in the particle plasma coupled to the scalar field sector responsible of the phase transition. Turbulent dynamics may also ensue in the plasma. Multiple GW production channels are then expected in a 1stO-PhT, in first place via bubble collisions~\cite{Witten:1984rs,Hogan:1986qda,Kosowsky:1991ua,Huber:2008hg,Caprini:2009fx}, and then through sound waves and turbulent motions~\cite{Kosowsky:1992rz,Kosowsky:2001xp,Weir:2017wfa,Caprini:2018mtu, Caprini:2019egz}. 

While in dark sectors GW backgrounds from 1stO-PhTs can peak across a wide frequency range~\cite{Schwaller:2015tja, Hall:2019ank}, electroweak-scale 1stO-PhTs in extended Higgs sectors, generate GW backgrounds observable with LISA~\cite{Caprini:2015zlo,Caprini:2019egz}. The latter scenarios can also be tied to baryogenesis mechanisms and dark matter candidates, so a connection emerges naturally between GW observations and BSM programs at the Large Hadron Collider (LHC) and future colliders ~\cite{Mazumdar:2018dfl,Hindmarsh:2020hop,Friedrich:2022cak}. There is, in general, great interest to study BSM scenarios predicting strong 1stO-PhTs and leading to sizable GW backgrounds within the reach of multiple detectors. Predicting the GW background spectrum is, however, a complicated and technically challenging task. To begin with, one needs to incorporate relativistic fluid dynamics~\cite{PencilCode:2020eyn,Brandenburg:2001wi,Brandenburg:2001yb}, in order to describe the particle plasma coupled to the scalar field responsible for the 1stO-PhT~\cite{Espinosa:2010hh,Hindmarsh:2013xza,Hindmarsh:2015qta,Hindmarsh:2017gnf}. As a consequence, intrinsic non-linearities in the fluid dynamics are eventually expected to become relevant, possibly leading to shocks in the sound waves, as well as to vorticity and turbulence~\cite{WilsonMatthews,Brandenburg:2001wi,Brandenburg:2001yb}. The shape of the SGWB spectrum is best understood for bubble collisions and acoustic production in certain regimes. From the turbulent stage, however, it is far less well understood~\cite{Pol:2019yex,Cutting:2019zws,Gogoberidze:2007an,Caprini:2009fx,Niksa:2018ofa,RoperPol:2021xnd,RoperPol:2022iel,Auclair:2022jod}, as simulations of turbulent flows are very challenging, whereas analytical calculations rely on assumptions that may require further testing. 

Only specialized numerical lattice simulations, typically with very high resolution and widely separated scales, will be able to tackle in full generality the production of GWs by 1stO-PhT's. Given the great interest in the topic, we plan to release a {\it MHD-Fluid} module for \CLns, dedicated to simulate the relativistic dynamics of fluids representing the plasma of particles coupled to the scalar sector responsible for a phase transition. This will include also the ability to solve magneto hydrodynamical (MHD) effects in situation where gauge fields participate in the dynamics, similarly as in the Pencil Code~\cite{PencilCode:2020eyn}. Furthermore, we expect to be able to set the initial conditions for the fluid dynamics as coming from proper phase transition dynamics involving a gauged Higgs sector coupled to Abelian and non-Abelian gauge fields.

Our starting point to describe the fluid dynamics is to consider the stress-energy tensor of a perfect fluid, $T^{\mu\nu} = (p + \rho) U^\mu U^\nu - p g^{\mu\nu}$, where $p$ and $\rho$ are the fluid's pressure and energy density, and we have introduced a 4-velocity as\footnote{Note that the $1/a$ factor included in $U^i$ ensures the desired normalization condition
$U^\mu U_\mu = U^\mu U^\nu g_{\mu\nu} = U^0 U^0 g_{00} + U^i U^j g_{ij} = \gamma^2 - \gamma^2 u^2 = -1$.} $U^\mu = \gamma (1, u^i/a)$, with standard relativistic $\gamma$-factor $\gamma \equiv 1/\sqrt{1 - u^2}$. The different components of the stress-energy tensor can then be written explicitly as
\begin{eqnarray}
    T^{00} &=& (p + \rho) \gamma^2 - p\,,\\
    T^{0i} &=& (p + \rho) \gamma^2 u^i/a\,,\\
    T^{ij} &=& (p + \rho) \gamma^2 u^i u^j/a^2 + p \delta^{ij}/a^2\, .
\end{eqnarray}

If we apply now the conservation of the energy momentum tensor in a curved background~\cite{Misner:1973prb,Brandenburg:1996fc}, we obtain
\begin{eqnarray}
D_\nu T^{\mu\nu} = \partial_\nu T^{\mu\nu} + \Gamma^\mu_{\sigma\nu}
    T^{\sigma\nu} + \Gamma^\nu_{\nu\sigma} T^{\mu\sigma}
    = 0,
    \label{eq:cons}
\end{eqnarray}
where $D_\mu$ is the gravitational covariant derivative. For a fluid with ultrarelativistic equation of state, with $p = \rho/3$, it is convenient to re-scale the components of $T^{\mu\nu}$ as $\tilde T^{00} = a^4 T^{00}$, $\tilde T^{0i} = a^{5} T^{0i}$, and
$\tilde T^{ij} = a^6 T^{ij}$. This allows to write the continuity and momentum equations in conformal time as~\cite{Brandenburg:1996fc,Jedamzik:1996wp}
\begin{eqnarray}
&&\partial_\eta \tilde T^{00} + \partial_i \tilde T^{0i} = 0\,,\label{eq:Fluid_EQ1}\\
&& \partial_\eta \tilde T^{0i} + \partial_j \tilde T^{ij} = 0\,.\label{eq:Fluid_EQ2}
\end{eqnarray}
As it turns out that it is possible to obtain $\tilde T_f^{ij}$ as a function of $\tilde T^{00}$ and
$\tilde T^{0i}$~\cite{Brandenburg:1996fc}, we can then evolve $\tilde T^{00}$ and $\tilde T^{0i}$ by solving Eqs.~(\ref{eq:Fluid_EQ1})-(\ref{eq:Fluid_EQ2}). Finally, one can always reconstruct the fluid variables $\rho$ and $u^i$ as a function of $\tilde T^{00}$ and
$\tilde T^{0i}$. 

For MHD effects (i.e.~gauge field interaction with the fluid), as well as for scalar-fluid coupling, one needs to specify a source term $\tilde S^\mu$ in the $rhs$ of Eqs.~(\ref{eq:Fluid_EQ1})-(\ref{eq:Fluid_EQ2}), which depends on the scalar $\phi$ and gauge field $A_\mu$ 
configurations, as well as on the spatial-spatial energy momentum component $\tilde T^{ij}$. We can then solve the system of fluid equations
\begin{eqnarray}
&&\partial_\eta \tilde T^{00} + \partial_i \tilde T^{0i} = \tilde S^0[\phi, \lbrace A_\mu \rbrace,\lbrace \tilde T_{lk} \rbrace]\,,\label{eq:FluidSource_EQ1}\\
&& \partial_\eta \tilde T^{0i} + \partial_j \tilde T^{ij} = \tilde S^i[\phi, \lbrace A_\mu \rbrace,\lbrace \tilde T_{lk} \rbrace]\,,\label{eq:FluidSource_EQ2}
\end{eqnarray}
together with the equations of motion describing the dynamics of $\phi$ and $A_\mu$, see Sects.~\ref{sec:scalar}-\ref{subsec:NonAbelian}, which now need to incorporate the fluid-coupling contributions as well. We are currently testing various implementation of fluid dynamics in \CLns. 

\section{Future update of \CLns, Part II: new features}  \label{sec:futurep2}

\subsection{Evolvers for non-canonical dynamics}
\label{sec:NonCanonicalDynamics}

The equations that describe the dynamics of interacting fields propagating in an expanding background, always have a common structure (assuming standard interactions leading to second order EOM), independently of the nature of the fields involved. Considering a certain set of fields $\lbrace f_{j}\rbrace$ and their conjugate momenta $\lbrace \pi_{j}\rbrace$, with $j$ labeling each degree of freedom (let them be e.g.~scalars, gauge field or gravitational wave components), we can always write these equations as
\begin{eqnarray}\label{eq:SchemeContVirgin1}
\pi_a(\eta) &=& a'(\eta)\,,\\
\label{eq:SchemeContVirgin2}
\pi_a'(\eta) &=& \mathcal{K}_a[a(\eta), E_V(\eta), E_K(\eta), E_G(\eta)]\,,\\
\label{eq:SchemeContVirgin3}
\pi_i({\bf x},\eta) &=& \mathcal{D}_i[f_i'({\bf x},\eta),a(\eta),\pi_a(\eta);\lbrace f_{j}({\bf x},\eta) \rbrace, \lbrace f'_{j\neq i}({\bf x},\eta) \rbrace]\,,\\
\label{eq:SchemeContVirgin4}
\pi_i'({\bf x},\eta) &=& \mathcal{K}_i[f_i({\bf x},\eta),\pi_i({\bf x},\eta),a(\eta),\pi_a(\eta);\lbrace f_{j\neq i}({\bf x},\eta) \rbrace, \lbrace \pi_{j\neq i}({\bf x},\eta) \rbrace]\,,
\end{eqnarray}
with the {\it drift $\mathcal{D}_i[...]$} defining the conjugate momentum of the $i$th degree of freedom ($dof$), and the {\it kernel} or {\it kick} $\mathcal{K}_i[...]$ determining the interactions of the $i$th $dof$ with the rest of $dof's$ (possibly including itself). If the expansion of the universe is sourced by the fields themselves, one also needs to specify the kick of the scale factor $\mathcal{K}_a[...]$, which represents the $rhs$ of Eq.~(\ref{eq:sfEOM}), i.e.~$\mathcal{K}_a[...]$ is sourced by the volume averages $\langle ... \rangle$ of the potential and the kinetic and gradient energy densities of the different $dof$'s. 

The current version of \CL provides a variety of evolvers for singlet-scalar and gauge-scalar theories, including $\mathcal{O}(\delta\eta^2)$ algorithms such as staggered leapfrog and velocity verlet, and $\mathcal{O}(\delta\eta^4)$-$\mathcal{O}(\delta\eta^{10})$ Yoshida algorithms based on velocity verlet. However, the implemented  algorithms are only envisaged for canonical field interactions, where the Kernels do not depend on conjugate momenta, i.e.~$\mathcal{K}_i({\bf x},\eta) \equiv \mathcal{K}_i[\lbrace f_{j}({\bf x},\eta)\rbrace,a(\eta)]$. When the kernel of a given $dof$ depends on conjugate momenta (typically on its own conjugate momentum\footnote{In such dependence we do not include the standard {\it friction} term $3\mathcal{H}\phi_i'$, which can be easily re-absorbed in appropriately defined conjugate momenta.}), the currently implemented methods do not represent good evolvers, and should not be used. As a matter of fact, some of the new physics cases presented in Sect.~\ref{sec:futurep1} exhibit kernels that depend on conjugate momenta, see e.g.~the $r.h.s.$ of the EOM of an axion and a gauge field in Eq.~(\ref{eq:AxionInfEOM}), or of a gravitationally non-minimally coupled scalar field in Eq.~(\ref{eq:EOMpiNMC}). In order to solve the EOM of those systems, we cannot use the default methods implemented in \CLns. We rather need to introduce {\it non-symplectic} methods, which can deal with kernels that depend on conjugate momentum $\mathcal{K}_i({\bf x},\eta) \equiv \mathcal{K}_i[\lbrace f_{j}({\bf x},\eta)\rbrace,\mathcal{K}_i[\lbrace \pi_{j}({\bf x},\eta)\rbrace,\pi_a(\eta),a(\eta)]$, such as e.g.~\textit{Runge-Kutta} or \textit{Gauss-Legendre}. While these methods can handle such kernels, they come with the disadvantage that they need extra memory to save auxiliar field configurations required in intermediate steps. Using these methods is however a `must-do' if one wants to solve field dynamics with non canonical interactions. We plan to update \CL with the addition of various flavours of such algorithms [considering at least an accuracy of $\mathcal{O}(\delta\eta^2)$, $\mathcal{O}(\delta\eta^{3})$ and $\mathcal{O}(\delta\eta^{4})$], adapted for specific problems such as axion-gauge dynamics, non-minimal gravitationally coupled scalars, fluid dynamics, non-canonical kinetic theories, and others.

\subsection{Arbitrary initial (spectrum) conditions}
\label{sec:arbitraryspec}

The default initial conditions currently implemented in \CLns, based on an initial spectrum of quantum fluctuations~\eqref{eq:QuantumFlucts}, may not be appropriate in some cases. In certain occasions, the user might want to provide a different initial spectrum. More specifically, given the definition of scalar power spectra, 
\begin{eqnarray}
\left\langle {\delta \phi}_{\bf k}{\delta \phi}_{{\bf k}'}^* \right\rangle \equiv (2\pi)^3 \mathcal{P}_{\delta \phi} (k)\delta(\bf{k}-\bf{k}') \ , 
\label{eq:ICVarPhi}\\ 
\left\langle {\delta \pi}_{\bf k}{\delta \pi}_{{\bf k}'}^* \right\rangle \equiv (2\pi)^3 \mathcal{P}_{\delta \pi} (k)\delta(\bf{k}-\bf{k}') \ , \label{eq:ICVar}
\end{eqnarray}
we will update \CL so that the user can have the option to provide an input text file specifying the power spectra functions $\mathcal{P}_{\delta \phi} (k), \mathcal{P}_{\delta \pi} (k)$, for each scalar field. The initial fluctuations will then be computed by setting the Fourier modes of the fields as 
\begin{eqnarray}
\delta  \phi ({  \bf k}) &= \frac{1}{\sqrt{2}} (|\delta  \phi^{l} ({  \bf k})|  e^{i \theta^{(l)} ({   \bf {k}}) } + |\delta  \phi^{r}({   \bf {k}})| e^{i \theta^{(r)} ({   \bf {k}}) }   ) \label{eq:fpr_influctbis} \ , \\
  \delta  \pi ({  \bf k}) &= \frac{1}{\sqrt{2}} (|\delta  \pi^{l} ({  \bf k})|  e^{i \theta^{(l)} ({   \bf {k}}) } + |\delta  \pi^{r}({   \bf {k}})| e^{i \theta^{(r)} ({   \bf {k}}) }   ) \label{eq:fpr_influctmom} \ , 
\end{eqnarray}
with $|\delta  \phi^{l,r} ({  \bf k})|$ and  $|\delta  \pi^{l,r} ({  \bf k})|$ random Gaussian fields with variances $\sigma_{\delta\phi}^2 =  \mathcal{P}_{\delta \phi} ({\bf k})$, $\sigma_{\delta\pi}^2 =  \mathcal{P}_{\delta \pi} ({\bf k})$, respectively. This feature will be available in an upcoming release of \CLns.

\subsection{Lattice simulations in $2+1$ dimensions}

We are typically interested in simulating field dynamics in 3+1 dimensions. However, there are scenarios where very long simulation times are required, and parallelization will not provide a wide enough dynamical range to probe all relevant physical scales. A possible way out might be to simulate the field dynamics in 2+1 dimensions, as long as this captures well the physics of 3+1 dimensions, which can only be assessed in a case by case basis. This trick reduces the simulation time by a factor $N$ (typically ${N \sim 10^2 - 10^3}$). More specifically, one could solve e.g.~the equations of a scalar field sector in a two-dimensional spatial layer, but still assuming that the spacetime metric is described by the FLRW metric~\eqref{eq:FLRWmetric} in 3+1 dimensions, with the scale factor evolution still given by the Friedmann equations~\eqref{eq:sfEOM}.

This technique has been used in Refs.~\cite{Antusch:2020iyq,Antusch:2021aiw,Antusch:2022mqv}, in scenarios where the inflaton oscillates around a monomial potential and broad parametric resonance of `daughter' fields coupled to the inflaton is developed. This allows e.g.~to characterize the long term evolution of the equation of state and field energy distribution after inflation. An explicit comparison between 2+1 and 3+1 lattice simulations is presented in the Appendix of \cite{Antusch:2020iyq}. There it is shown that the equation of state and the field spectra evolve almost identically in 2+1 and 3+1 dimensions, demonstrating the validity of this {\it dimensional reduction} idea for broad parametric resonance.

We note that while the current version of \CL is already capable of solving dynamics in 2+1 dimensions, it is still necessary to adapt the {\tt CosmoInterface} library for this purpose for specific problems. We plan to release an updated version of \CL with the capability of evolving scalar field sectors in $2+1$ dimensions in the near future. 

\subsection{Discrete spatial derivatives of higher accuracy}

In order to solve the field equations in the lattice, we replace the continuous spatial derivatives by finite difference formulas that approximate them to the continuum to an accuracy of order $\mathcal{O}(\delta x^m)$, with $m \geq 1$. For example, in the case of a scalar field sector, first spatial derivatives appear in the gradient energy contribution to the Friedmann equation \eqref{eq:pLocal}, while second derivatives appear in the scalar equation of motion \eqref{eq:EOMscalarContinuumNat}.

Let us consider a continuous function ${\tt f} ({\bf x})$ and its lattice representation $f({\bf n})$, where the vector ${\bf n} = (n_1, n_2, n_3)$ (with $n_i = 0, \dots N-1$ and $i=1,2,3$) tags the sites of a 3-dimensional lattice. The first derivative at the lattice site ${\bf x} \equiv {\bf n} \dx$ in the $i$-spatial direction can be approximated, up to second order of accuracy, by the following \textit{centered} finite difference,
\begin{eqnarray}
    \label{eq:neutrald}
    [\nabla^{(0)}_i f] = \frac{f({\bf n}+ \hat{\imath}) - f({\bf n}- \hat{\imath})}{2\dx} ~~\longrightarrow ~~ \partial_i{\tt f}({\bf x})\big|_{{\bf x}\,\equiv\, {\bf n}\dx} + \mathcal{O}(\dx^2)\,,
\end{eqnarray}
where $ \hat{\imath}$ are unit vectors in the $i$-spatial direction of the lattice (corresponding to positive displacements of length $\dx$). A disadvantage of such construction  is that it is insensitive to field variations in the smallest possible distance $\dx$. Alternatively, we can define the following \textit{charged} forward/backward derivatives,
\begin{eqnarray}
    \label{eq:forwardbackwardd}
    [\nabla^\pm_i f] = \frac{\pm f({\bf n}\pm \hat{\imath}) \mp f({\bf n})}{\dx} ~~\longrightarrow ~~ \left\lbrace\begin{array}{l}
        \partial_i{\tt f}({\bf x})\big|_{{\bf x}\,\equiv\, {\bf n}\dx} + \mathcal{O}(\dx)\ ,  \vspace*{0.2cm}\\
        \partial_i{\tt f}({\bf x})\big|_{{\bf x}\,\equiv\, ({\bf n} \pm \hat\imath/2)\dx} + \mathcal{O}(\dx^2) \ ,
    \end{array}\right.
\end{eqnarray}
which approximate the continuous derivative at grid points to only first order of accuracy, but approximates the derivative at `half-way' points to second order. In the first case we have a \textit{one-sided} (first-order accurate) approximation, while in the second case we have a \textit{centered} (second-order accurate) approximation. Similarly, we can build a finite difference expression for second derivatives as follows,
\begin{eqnarray}
    \label{eq:laplaciand}
 \hspace*{-2.4cm}   [\nabla^2_i f] \equiv [\nabla^-_i \nabla^+_i f] =  \frac{f({\bf n}+ \hat{\imath}) - 2 f({\bf n}) + f({\bf n} - \hat{\imath}) }{\dx^2} ~~\longrightarrow ~~ \partial_i^2{\tt f}({\bf x})\big|_{{\bf x}\,\equiv\, {\bf n}\dx} + \mathcal{O}(\dx^2)
\end{eqnarray}
which corresponds to a second-order accurate, centered approximation at a grid point.

The current version of \CL uses Eqs.~\eqref{eq:forwardbackwardd} and \eqref{eq:laplaciand} to approximate the spatial derivatives in the Friedmann and field equations respectively. However, there are scenarios where one may want to use derivatives of higher order accuracy. For example, fourth order-accurate spatial derivatives have been used in \cite{Antusch:2016con} in order to minimize discretization errors in the simulation of gravitational waves from post-inflationary oscillon dynamics. 

In Ref.~\cite{DiscreteDerivatives}, a recursive method is developed that obtains finite difference expressions for any order of derivative and accuracy. The weight that each lattice site contributes to the finite difference can be obtained through a recursive formula. This way, tables are obtained for derivatives at both grid points and `half-way' points, including both centered and one-sided approximations. For illustrative purposes we provide, for the first and second spatial derivatives, centered finite differences at grid points that are accurate to fourth order,
\begin{eqnarray}
  \hspace*{-1cm}  [\nabla_i f]^{(4)} &\equiv \frac{f({\bf n}-2 \hat{\imath}) - 8 f ({\bf n}- \hat{\imath}) +8 f({\bf n}+ \hat{\imath})  - f ({\bf n}+2 \hat{\imath})}{12\, \dx}  \label{eq:der4a} \\[7pt]
   & ~~\longrightarrow ~~ \partial_i{\tt f}({\bf x})\big|_{{\bf x}\,\equiv\, {\bf n}\dx} + \mathcal{O}(\dx^4) \ ,  \nonumber \\[7pt]
     \hspace*{-1cm} [\nabla_i^2 f]^{(4)} & \equiv  \,\frac{- f ({\bf n}+2  \hat{\imath}) + 16 f ({\bf n}+ \hat{\imath})  -30 f({\bf n} ) +16 f({\bf n}- \hat{\imath}) - f ({\bf n}-2 \hat{\imath})}{12\,\dx^2}  \label{eq:der4b} \\[7pt]
    & ~~\longrightarrow ~~ \partial_i^2{\tt f}({\bf x})\big|_{{\bf x}\,\equiv\, {\bf n}\dx} + \mathcal{O}(\dx^4)  \nonumber \ .  
\end{eqnarray}

Note that computing spatial derivatives in parallelized simulations is challenging because each  processor only has access to a subset of the entire lattice. This means that, when computing spatial derivatives at the boundaries of each sublattice, information must be passed from other processors using e.g.~MPI. In \CL this is dealt by introducing extra \textit{ghost} layers at the boundaries to store this information, see Ref.~\cite{Figueroa:2021yhd} for more details. The width of these layers is controlled by the \textit{ghost cells} parameter, which can be easily modified at compilation time in the code. The spatial derivatives \eqref{eq:forwardbackwardd}-\eqref{eq:laplaciand} require only one ghost cell, but \eqref{eq:der4a}-\eqref{eq:der4b} require two. Note that increasing the number of ghost cells inevitably makes the code less scalable and hence slower. We plan to implement some higher order accuracy spatial derivatives in a future version of \CLns.

\section{Final outlook} \label{sec:outlook}

Since its public release in February 2021, \CL has been used to explore different aspects of the non-linear dynamics of the early universe, including: (p)reheating after inflation \cite{Antusch:2020iyq,Antusch:2021aiw,Antusch:2022mqv,Dux:2022kuk,Garcia:2023eol,Garcia:2023dyf,Matsui:2023wxm,Matsui:2023ezh,Mansfield:2023sqp}, the impact of such era on inflationary CMB observables \cite{Antusch:2020iyq,Antusch:2021aiw,Mansfield:2023sqp}, the generation of a relic density of dark matter \cite{Garcia:2021iag,Lebedev:2022ljz,Garcia:2022vwm,Lebedev:2022vwf,Zhang:2023xcd,Zhang:2023hjk}, the production of primordial gravitational waves from oscillating scalar fields \cite{Figueroa:2022iho,Cosme:2022htl,Cui:2023fbg,Mansfield:2023sqp,Garcia:2023eol}, the study of scalar theories with non-minimal gravitational interactions~\cite{Figueroa:2021iwm,Lebedev:2022vwf,Lebedev:2023zgw,Laverda:2023uqv,Figueroa:2024asq} and axion-gauge interactions~\cite{Figueroa:2023oxc}, cosmic defects and associated gravitational wave production~\cite{Baeza-Ballesteros:2023say,Li:2023gil,Ramberg:2022irf}, phase transitions~\cite{Li:2023yzq}, and oscillons~\cite{Mahbub:2023faw,Piani:2023aof}.

We plan to publicly release {\tt version 2.0} of \CL in the foreseeable future, including, among other new physics modules, the capability of simulating axion-gauge interactions, non-minimal gravitational scalar interactions, and the creation and evolution of cosmic defect networks, as described in  Sect.~\ref{sec:futurep1}. We plan to incorporate as well some of the technical features described in Sect.~\ref{sec:futurep2}. We expect that with these and future updates, \CL will continue being a relevant tool for the scientific community interested in $\mathcal{L}${\it attice} $\mathcal{C}${\it osmology} problems.

\section*{Acknowledgments}

We are very grateful to Wessel Valkenburg, who collaborated with us in the original creation of the code. We are also very thankful to Jorge Baeza-Ballesteros, Joanes Lizarraga, Nicolas Loayza, Kenneth Marschall, Antonino Midiri, Toby Opferkuch, Alberto Roper Pol, Ben A. Stefanek and Ander Urio, for collaborating with us and contributing to further developing the code. We are also thankful to the participants of the 2022 and 2023 editions of the \CL schools, which took place at Valencia and on-line respectively. DGF (ORCID 0000-0002-4005-8915) was supported by a Ram\'on y Cajal contract with Ref.~RYC-2017-23493 during completion of this work, and by a Generalitat Valenciana grant PROMETEO/2021/083, and Spanish Ministerio de Ciencia e Innovacion grant PID2020-113644GB-I00. FT (ORCID 0000-0003-1883-8365) is supported by a \textit{Mar\'ia Zambrano} fellowship (UV-ZA21-034) from the Spanish Ministry of Universities and grant PID2020‐116567GB‐C21 of the Spanish Ministry of Science.

\section*{References}

    \bibliography{biblio,extra}
    \bibliographystyle{ieeetr}

\end{document}